\documentstyle[12pt,aasms4]{article}

\input epsf
\newdimen\fighsize \def\epsscale#1{\fighsize=#1\hsize} \epsscale{1}
\def\plotone#1{\epsfxsize=\fighsize\centerline{\epsfbox{#1}}}
\def\plottwo#1#2{\centerline{\epsfxsize=.5\fighsize\epsfbox{#1}\hss
                             \epsfxsize=.5\fighsize\epsfbox{#2}}}
\def\plotfour#1#2#3#4{\centerline{\epsfxsize=.5\fighsize\epsfbox{#1}\hss
                                  \epsfxsize=.5\fighsize\epsfbox{#2}}
                      \centerline{\epsfxsize=.5\fighsize\epsfbox{#3}\hss
                                  \epsfxsize=.5\fighsize\epsfbox{#4}}}

\begin{document}

\font\bmit=cmmib10
\def\vtheta{\hbox{\bmit\char"12}}
\def\vthes{\hbox{\bmit\char"12}_{\rm s}}
\def\Dfac{{D_{\rm ol}D_{\rm os}\over D_{\rm ls}}}
\def\zl{z_{\rm l}}
\def\<#1>{\langle\hbox{#1}\rangle}
\def\half{\hbox{$1\over2$}}
\def\onethird{\hbox{$1\over3$}}  
\def\twothirds{\hbox{$2\over3$}}

\title{Pixelated Lenses and $H_0$ from Time-delay QSOs}

       \author{Liliya L.R. Williams}
       \affil{Department of Physics and Astronomy\\  
              University of Victoria\\
              Victoria, BC, V8P 1A1, Canada}
       \and
       \author{Prasenjit Saha\altaffilmark{1}}
       \affil{Department of Physics\\  
	      Oxford University\\
              Keble Road,\\
  	      Oxford, OX1 3RH, UK}

\altaffiltext{1}{Present address: Astronomy Unit, School of
Mathematical Sciences, Queen Mary and Westfield College, London
E1~4NS, UK}

\begin{abstract}
Observed time delays between images of a
lensed QSO lead to the determination of the Hubble constant by
Refsdal's method, provided the mass distribution in the lensing galaxy
is reasonably well known.  Since the two or four QSO images usually
observed are woefully inadequate by themselves to provide a unique
reconstruction of the galaxy mass, most previous reconstructions have
been limited to simple parameterized models, which may lead to large
systematic errors in the derived $H_0$ by failing to consider enough
possibilities for the mass distribution of the lens. We use
non-parametric modeling of galaxy lenses to better explore physically
plausible but not overly constrained galaxy mass maps, all of which
reproduce the lensing observables exactly, and derive the
corresponding distribution of $H_0$'s.  Blind tests---where one of us
simulated galaxy lenses, lensing observables, and a value for $H_0$,
and the other applied our modeling technique to estimate $H_0$
indicate that our procedure is reliable. For four simulated lensed
QSOs the distribution of inferred $H_0$ have an uncertainty of
$\simeq10\%$ at 90$\%$ confidence.  Application to published
observations of the two best constrained time-delay lenses, PG1115+080
and B1608+656, yields $H_0=61\pm11\rm\,km\,sec^{-1}\,Mpc^{-1}$ at 68\%
confidence and $61\pm18\rm\,km\,sec^{-1}\,Mpc^{-1}$ at 90\%
confidence.
\end{abstract}

\section{Introduction}

Most ways of measuring the Hubble constant involve a form of distance
ladder, which utilizes a number of astrophysical standard candle and
standard ruler relations, and is calibrated locally by a geometrical
technique such as parallax
(e.g., Madore et al.\ 1999, Madore et al.\ 1998, Kennicutt 1995). 
A recent exciting development in this field is
to extend the reach of the geometrical rung of the distance ladder by
using masers in orbit around galaxy centers to get distances to nearby
galaxies thus bypassing Cepheids (Herrnstein et al.\ 1999). A few
methods involve no distance ladder: good examples are (i)~inferring
the distance of Type~II supernovae from their light curves and spectra
by modeling their expanding photospheres (Schmidt et al.\ 1992),
and (ii)~comparing the $H_0$-independent angular extent of galaxy
clusters to their $H_0$-dependent depth as deduced by the X-ray
emission, and the Sunyaev-Zeldovich microwave background decrement due
to the cluster (Hughes \& Birkinshaw 1998).

But the most `one-step' method of all was proposed by S. Refsdal in 1964,
though it has only recently become feasible.  The principle of
Refsdal's method is simple.  In a system where a high-redshift QSO is
split into multiple images by an intervening galaxy lens, the
difference in light travel time between different images (observable
as time delays if the QSO is variable) is proportional to the scale
factor of the universe.  The time delay is given by the schematic
formula
\begin{eqnarray}
\<Time delay> & = & h^{-1} \times \<1 month> \times
\<image separation in arcsec>^2 \nonumber\\
& & \times z_{\rm lens}\times\<weak dependence on $z_{\rm lens}$,$z_{\rm QSO}$,
and cosmology> \\
& & \times \<lens-mass-distribution dependent factor> \nonumber
\label{schem_eq}
\end{eqnarray}
where the last two factors are of order unity.  To obtain $H_0$ using
this method one requires three types of input: (i)~the observed time
delay(s) between QSO images, (ii)~knowledge of the cosmology, and
(iii)~the mass distribution in the lensing galaxy.  The first can and
has been measured with increasing precision for about eight systems so
far.  The second is not a serious problem, because the dependence on
cosmology is weak and the uncertainty due to it is easy to quantify;
in this paper we will refer all results to the Einstein-de Sitter
cosmology.  The uncertainty in $H_0$ is dominated by the third item;
the number of usable constraints on the mass distribution in the
galaxy are few, while the range of possible distributions is
huge. Thus, mass distribution is the major source of uncertainty.

Two different paths can be taken to compensate for our lack of
knowledge about the galaxy. One is to assume an exact parametric form
for the galaxy mass distribution and fit the observed lensing
properties as best as possible; the other is to take the image
properties as exact, and try to reconstruct the galaxy mass map as
best as possible.  Single parametric models fix the last term in
(\ref{schem_eq}) and thus cannot account for the uncertainty resulting
from it.  Blandford \& Kundi\'c (1996) advise that even if one finds a
parametric galaxy model which is dynamically possible and which
reproduces the image properties with acceptably low $\chi^2$, one
still has to `aggressively explore all other classes of models' to get
the true uncertainty in $H_0$. To explore the model space in a
systematic fashion one needs to use a representation of the galaxy
mass distribution that is general and not restricted to a particular
form. One way would be to expand the mass distribution using a set of
basis functions, another is to pixelate the galaxy and take each
pixel as an independent mass element. We introduced pixelated models
in Saha \& Williams (1997, hereafter SW97) but at that time did not
have any strategy for searching model space.  We have now extended
that work to explore the model space with the goal of estimating the
uncertainty in the derived value of $H_0$.

The plan of this paper is as follows. In Section~\ref{obs} we
summarize the observational situation with regard to strongly lensed
QSOs.  In Section~\ref{formalism} we present the general lensing
formalism and point out a few properties of the lensing equations that
are useful in interpreting the results of modeling.  We also explain
the reasons for confining our analysis to PG1115+080 and B1608+656 for
now.  Sections~\ref{method} and ~\ref{blind} describe our method for
deriving $H_0$ and test it on a synthetic sample via a blind
test. Application to the real systems can be found in
Section~\ref{real}. Section~\ref{summary} discusses our results.

\section{Observed Time-Delay Lenses}\label{obs}

The first piece of input for $H_0$ determination is the measurement of 
time delays between the various QSO images. At the present time,
ten multiply-imaged QSOs already have measured time
delays or are being monitored: 
Q0957+561 (Kundi\'c et al. 1997a), 
PG1115+080 (Schechter et al. 1997, Barkana 1997), 
B1608+656 (Fassnacht et al. 1999),
B0218+257 (Biggs et al. 1999), 
PKS 1830-211 (Lovell et al. 1998),
HE 1104-1805 (Wisotzki et al. 1998),
B1030+074, 
B1600+434 (Burud et al. 1999), 
J1933+503, and 
RXJ0911+0551 (Hjorth et al. 1999).
In this work we limit ourselves to 4-image lenses with known source and lens
redshifts and accurate time delay measurements; PG1115+080 and B1608+656 
fit the description.

PG1115 (Weymann et al. 1980) was the second lens to be discovered. The source 
is a radio-quiet QSO at $z_s=1.722$. Accurate positions for the images were 
measured by Kristian et al. (1993); lightcurves were analyzed by Schechter 
et al. (1997), and time delays derived by Schechter et al. and Barkana (1997).
The main lensing galaxy is an outlying member of a small galaxy group, 
$z_l=0.311$ with an estimated line of sight velocity dispersion of 
$270\pm70$km~s$^{-1}$ (Kundi\'c et al. 1997b).  A summary of observational 
results on this system can be found in SW97. 

B1608 was discovered in the Cosmic Lens All-Sky Survey (Myers et al. 1995, 
Myers et al. 1999). 
The lens is either a perturbed single galaxy or a merging/interacting
pair of galaxies superimposed in the plane of the sky. The source and lens
redshifts are 1.394 and 0.630 respectively. The time delays were recently
reported by Fassnacht et al. (1999) based on VLA observations spanning 7 
months. The time delays we use in this work are an earlier determination
(Fassnacht, private communication), and are less than 0.5$\sigma$ away from
the values quoted in Fassnacht et al. (1999); 
$\Delta t_{BA}=28.5$, $\Delta t_{BC}=32$, and $\Delta t_{BD}=77$.

\section{Lensing formalism}\label{formalism}

A photon traveling through a galaxy will take longer to arrive at the
observer then an unimpeded photon. Part of the time delay occurs
because the path of the ray bundle makes a detour rather than going
straight; the time delay is further increased because the photon
travels through the gravitational potential well of the galaxy. The
total time delay is given by,
\begin{equation}
\tau(\vtheta,\vthes)=(1+\zl) \Dfac \left[\half(\vtheta-\vthes)^2-
{1\over \pi}\int\!d^2\vtheta'\,\kappa(\vtheta') \ln|\vtheta-\vtheta'|
\right]
\label{tau_eq}
\end{equation}
where $\vtheta$ is the position on the sky, $\vthes$ is the source
position, $D$'s are the angular diameter distances between the source,
the lens and the observer, $\zl$ is the redshift of the lens galaxy,
and $\kappa(\vtheta)$ is the projected mass density in the galaxy in
units of $\Sigma_{\rm crit}=(c^2/4\pi G)(D_{\rm os}/D_{\rm ls}D_{\rm
ol})$.

If the lens mass distribution $\kappa(\vtheta)$ is known then the
arrival time surface, Eq.\ (\ref{tau_eq}) provides us with all the
necessary information about the images. Time delay between any two
images is just the difference between $\tau$ at the relevant
locations. According to Fermat's Principle the images appear at
stationary points of the arrival time surface,
\begin{equation}
{\partial\tau\over\partial\vtheta} = 0 =
\vtheta-\vthes-{1\over\pi}\int\!d^2\vtheta'\kappa(\theta')
{\vtheta-\vtheta' \over |\vtheta-\vtheta'|^2}
\label{alpha_eq}
\end{equation}
Image distortion and magnification are given by the inverse of the curvature
matrix of the arrival time surface
\begin{equation}
\left[\partial^2\tau\over\partial\theta_i\theta_j\right]^{-1}
\end{equation}

\begin{figure}
\plotone{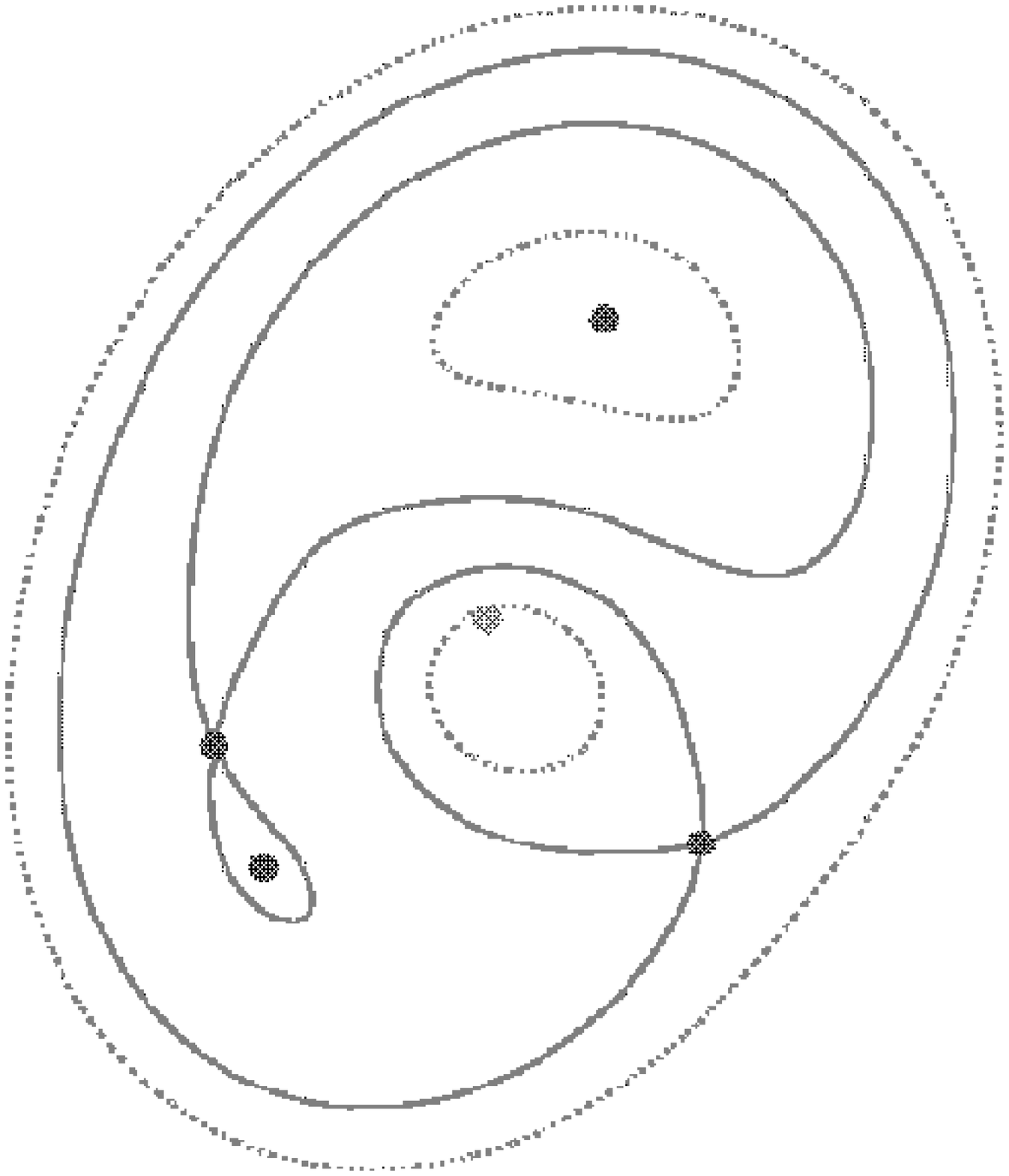}
\caption
{Generic arrival-time surface for a four-image QSO (in fact, a
possible one for PG1115).  The black filled circles mark the images
and the gray filled circle marks the source, while the curves are
isochronal contours.  This is only one of six possible configurations
for a 4+1 image system (Section 5.5 of Schneider et al.\ 1992), but
the most natural one where the lens is very roughly circular.  The top
image is the lowest minimum in the arrival-time surface and hence the
first image.  Next comes the minimum at lower left, followed closely
by the saddle point at left, and finally the saddle point at lower
right.  The contours indicate one more image---a maximum close to the
source---but this is not seen in real systems; it is presumed to be
demagnified below visibility because of the centrally peaked mass
distribution in lensing galaxies.
\label{morph_fig}}
\end{figure}

A few things can be learned by looking at the arrival time and lens 
equations:

(1) The time ordering of the images can be deduced from the image
configuration using the morphological properties of the arrival time
surface.  The image furthest from the lensing galaxy is always
first, and the one nearest the galaxy the last.  In four-image QSOs
the second image is the one opposite the first.  Figure \ref{morph_fig}
illustrates.

(2) When four images are formed by an isolated galaxy of typical
ellipticity the images are located nearly at the same galactocentric
distance. This is easy to see by considering the two pieces of the
arrival time surface. If the source and the center of the galaxy are
not well aligned, i.e., if the `bump' due to the gravitational
potential contribution is away from the `well' of the geometrical
contribution, then the steepness of the geometrical part allows only
two images to form, one roughly on either side of the galaxy
center. To get four images, the bump of the gravitational contribution
must be centered close to the source location. In such a situation the
total arrival time configuration is centrally symmetric and the
resulting images are approximately equidistant from the galaxy center.

(3) If the four images of a single source are located at different
galactocentric distances the simplest explanation is the presence of
external shear. External shear effectively raises the gravitational
part of the arrival time surface closest to itself (see Fig.\
\ref{morph_fig} and Eq.\ \ref{tau_eq}). The effect is to push the
locations of the stationary points away from the source of external
shear, hence increasing the radial spread of images. It follows that
the direction of external shear can be determined by examining image
locations with respect to the galaxy center. PG1115 is a good example;
the image closest to the galaxy center, image B, is located between
the galaxy-lens and the galaxy group, which is the source of external
shear in this case.

(4) Position angles (PA) of images are determined by the ellipticity PA of 
the galaxy roughly at the radius of the images. When images are spread 
over a range of radial distances their PA provide information on galaxy 
ellipticity PA over a range of galactocentric distances. Thus detailed
modeling can reveal the twisting of the isodensity contours. 

(5) Not all types of information about images are equally useful for
modeling purposes. The arrival time surface integrates over
$\kappa(\vtheta)$ twice, making time delays most sensitive to the
overall mass distribution in the galaxy, and least dependent on the
local small-scale perturbations in the mass distribution. Image
positions are determined from the lensing equation which integrates
over $\kappa(\vtheta)$ once. Finally, image magnifications are very
dependent on the local behavior of mass, making them the least useful
for modeling.  This means, unfortunately, that a double like
Q0957, though it has well-measured substructure in the images and
near-perfect time-delay measurements, provides too few constraints on
the lensing mass to usefully estimate $H_0$ unless drastic
assumptions about the mass distribution are made. In that case, the
derived errors will tend to be underestimated as was noted by Bernstein 
and Fischer (1999) who constructed many types of parametric models for 
Q0957: `The bounds on $H_0$ are strongly dependent on our assumptions 
about a ``reasonable'' galaxy profile'.

(6) A linear rescaling of the arrival time and lens equations, i.e.,
multiplying both by a constant factor $\epsilon$ will not alter the
observable properties of images, image separations and relative
magnification tensors. Physically the transformation amounts to
rescaling the mass density of the lens by $\epsilon$ and adding a
constant mass density sheet.  This transformation was first discussed
by Gorenstein et al.\ (1988) with regard to modeling of Q0957, and
later became known as the mass sheet degeneracy.  Note that a mass
sheet extending to infinity is not needed, a mass disk larger than the
observed field is enough because an external monopole has no
observable effect.

\section{The method}\label{method}

The first step is to pixelate the lens plane mass distribution of the
main lensing galaxy. In practice we use $\sim0.1''$ pixels, and limit
the galaxy to a circular window of radius about twice that of the
image-ring. Pixelated versions of Eqs.\ (\ref{tau_eq}) and
(\ref{alpha_eq}) are:
\begin{equation}
\tau(\vtheta,\vthes) = (1+\zl)\Dfac
\left[ \half|\vtheta|^2 - \vtheta\cdot\vthes
-\sum_n \kappa_n \psi_n(\vtheta) \right]
\label{pixtau_eq}
\end{equation}
and
\begin{equation}
\vtheta - \vthes - \sum_n \kappa_n\vec\alpha_n(\vtheta) = 0,
\label{pixalpha_eq}
\end{equation}
where the summation is over mass pixels and $\psi_n$ and $\alpha_n$
are integrals over individual pixels and can be evaluated analytically
(see Appendix of SW97). A term $|\vthes|^2$ has been omitted from Eq.\
(3) because a constant additive factor in the arrival time cannot be
measured.

Image properties translate into linear constraints in the
$(N\!+\!2)$-dimensional model space, where $N$ dimensions represent a
pixel each and 2 represent source coordinates. We call these primary
constraints. The images can provide us with only a few constraints: in
a 4-image system we have $2\times 4$ coordinates and 3 time delay
ratios: 11 in all.  On the other hand, the unknowns are numerous,
$\sim20^2$ mass pixels plus 2 source coordinates. This results in a
plethora of galaxy models each of which reproduces the image
properties exactly. Luckily, the bulk of these models can be discarded
because they do not look anything like galaxies. In fact, we consider
only those models which satisfy the following further (linear) constraints,
which we call secondary. These pertain to the main lensing galaxy:
\begin{enumerate}
\item mass pixel values, $\kappa_n$ must be non-negative;
\item the location of the galaxy center is assumed to be coincident
with that of the optical/IR image;
\item the density gradient of the lens must point no more
than $45^\circ$ away from the center of the galaxy;
\item the lens must have inversion symmetry, i.e., look the same if
rotated by $180^\circ$ [enforced only if the main lensing galaxy
appears unperturbed and has no companions close to the QSO images];
\item logarithmic projected density gradient in the vicinity of the
images, $d\ln\kappa/d\ln\theta={\rm ind}(r)$ should be no shallower that
$-0.5$. For a power law projected density profile, radial magnification
at an image is equal to $-1/{\rm ind}(r)$, therefore a statement that
${\rm ind}(r)<-0.5$ means that images are magnified radially by less than a
factor of 2, which is probably a reasonable limit given the appearance
of optical Einstein rings seen in some systems, for example, PG1115
and B0218;
\item external shear, i.e., influence of mass other than the main
lensing galaxy is restricted to be constant across the image region,
and is represented by adding a term
$\half\gamma_1(\theta_1^2-\theta_2^2)+\gamma_2\theta_1\theta_2$ to
the lensing potential, Eq.\ (\ref{tau_eq}).
\end{enumerate}

All these constraints are non-restrictive and are obeyed by the vast
majority of galaxies, thus our analysis explores the widest possible
range of galaxy mass distributions.

Obviously, the primary and secondary constraints are not enough to
isolate a unique galaxy mass solution. A unique solution can be
singled out by further specifying galaxy properties. For example, in
SW97 particular galaxy models were found making a trial value of $H_0$
as one of the primary constraints, and then and picking the model that
followed the observed light distribution as closely as possible given
the rigid primary and secondary constraints, see Figures 2--5 of
SW97. Here our aim is different.

Any of the infinitely many models remaining after the primary and
secondary constraints have been applied could be the real lens, as all
of them reproduce the image properties exactly and all look reasonably
like a galaxies, therefore any one of the corresponding derived
$H_0$'s could be the real $H_0$.  We want to produce an ensemble that
samples this model space, and our procedure is as follows.

The allowed models form a simplex in the $(N\!+\!2)$-dimensional space
of mass pixels and source positions, because the constraints are all
linear.  We start with a random point in the allowed simplex (i.e., an
allowed model).  Next we choose a random vertex of that simplex, which
is easily done by linear programming.  Then we consider the line
joining the current point with the vertex, and move to a random point
on it, taking care to remain inside the simplex.  The process is
repeated until a sample of 100 models, and hence 100 $H_0$ values, is
assembled.  This procedure is a trivial case of the Metropolis
algorithm (see e.g., Binney et al.\ 1992) for sampling density
functions in high-dimensional spaces.

The resulting ensemble of $H_0$ values has a straightforward
interpretation in terms of Bayesian probabilities.  The part of model
space allowed by the secondary constraints is the prior (i.e.,
possibilities allowed before considering tha data).  Our prior is
uniform, which is to say that we have not incorporated any prior
preferences between different models allowed by the secondary
constraints. Since the unknowns $\kappa_n$ occur linearly in Eqs.\
~\ref{pixtau_eq} and ~\ref{pixalpha_eq}, a uniform prior means that
any linear interval in $\kappa_n$ is a priori as probable as any other
interval of equal length. The primary constraints come from data, and
the 100 models that satisfy both primary and secondary constraints
sample the posterior probability distribution.  At the present time
there is no clear motivation to use any other but a uniform prior,
however, a non-uniform prior, if desired, would modify the method only
slightly: one could either keep the same 100 models but weight them
according to the prior, or take the prior into account while choosing
the models through the Metropolis prescription.

\section{Blind tests of the method}\label{blind}

Before applying the method to real systems we try it on a synthetic
situation designed to resemble the real world as close as possible.
One of us, ``Person A'', picked an $h$ value and created a set of four
galaxies and the corresponding images of a single background source in
each case. Exact values of image positions with respect to the galaxy
center and time delays (but not $h$, nor information as to whether the
galaxy was inversion symmetric or if there was any external shear)
were conveyed to the other one of us, ``Person B'', who used this
information to construct an ensemble of galaxy models and derive $h$
distributions for each case separately.

\begin{figure}
\plotone{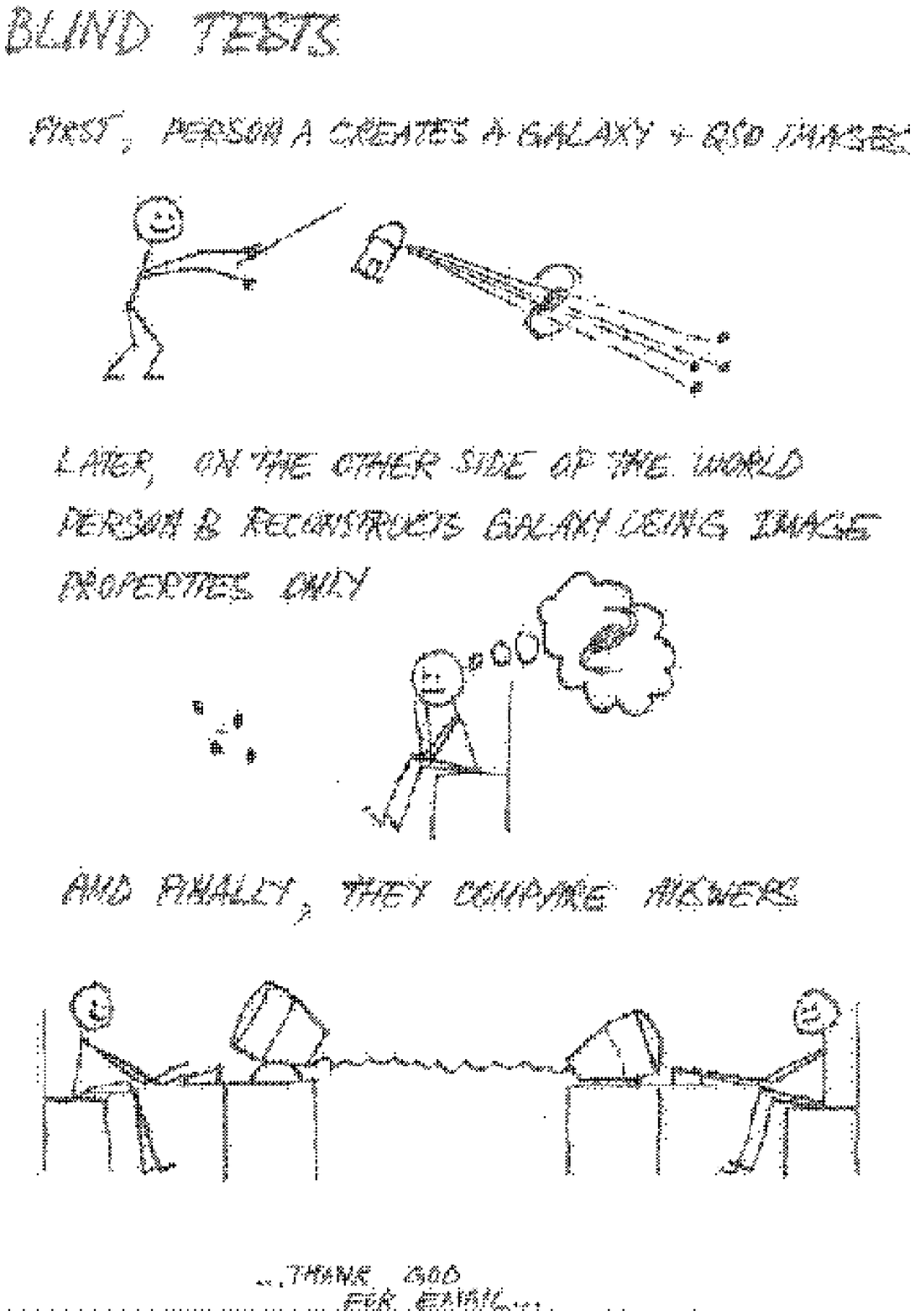}
\caption
{Pictorial illustration of the three stages of the blind tests 
designed to quantify the trustworthiness of our $H_0$ estimation method.
\label{cartoon_fig}}
\end{figure}

We ran the whole experiment several times to remove bugs, and
did not want to fall into the trap of simply publishing the results
of the best run.  So once we were confident that the experiment
worked, we decided that the next four galaxies, whatever the
results, would go into the published paper. Figure~\ref{cartoon_fig} 
pictorially illustrates the three stages of our blind test.

\begin{figure}
\plottwo{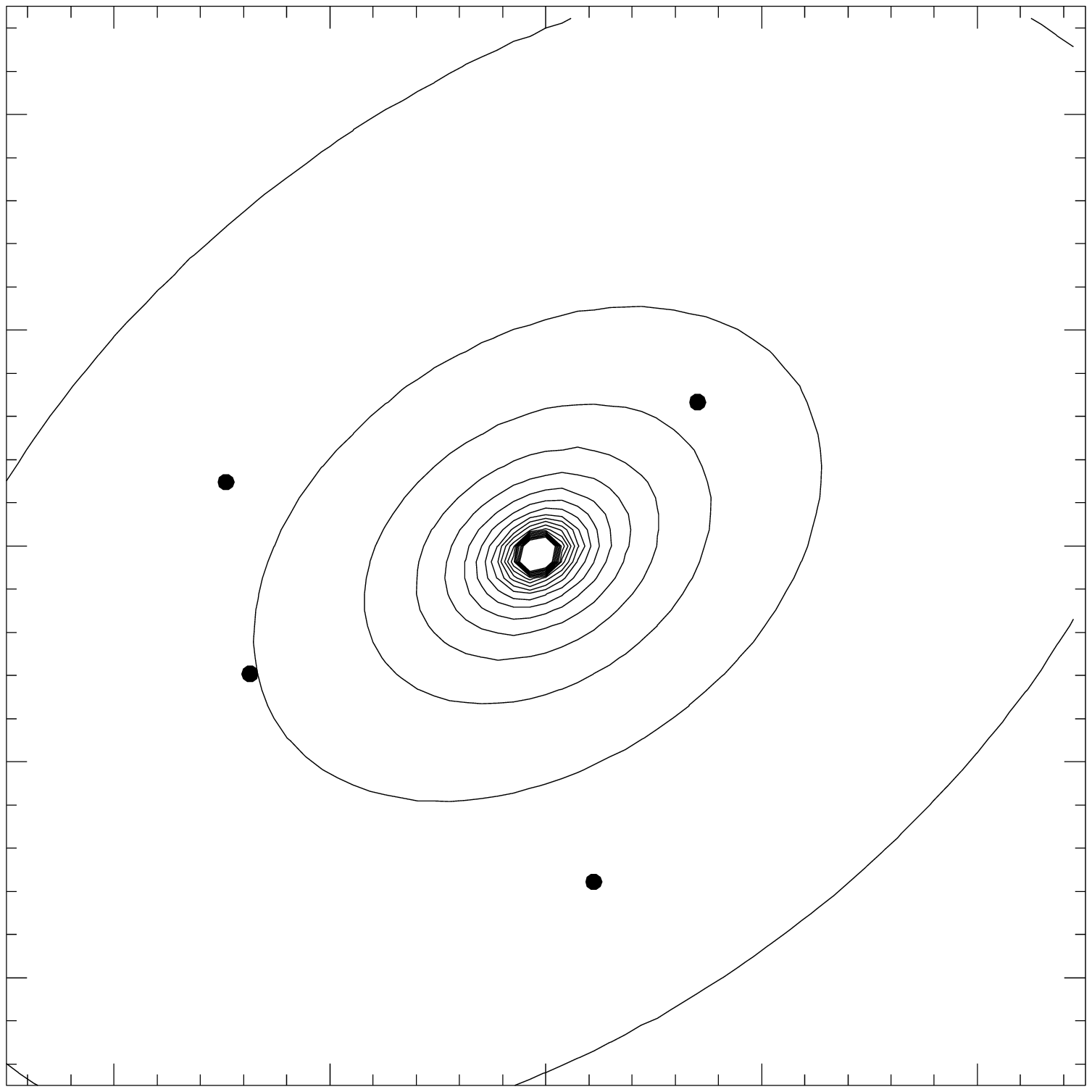}{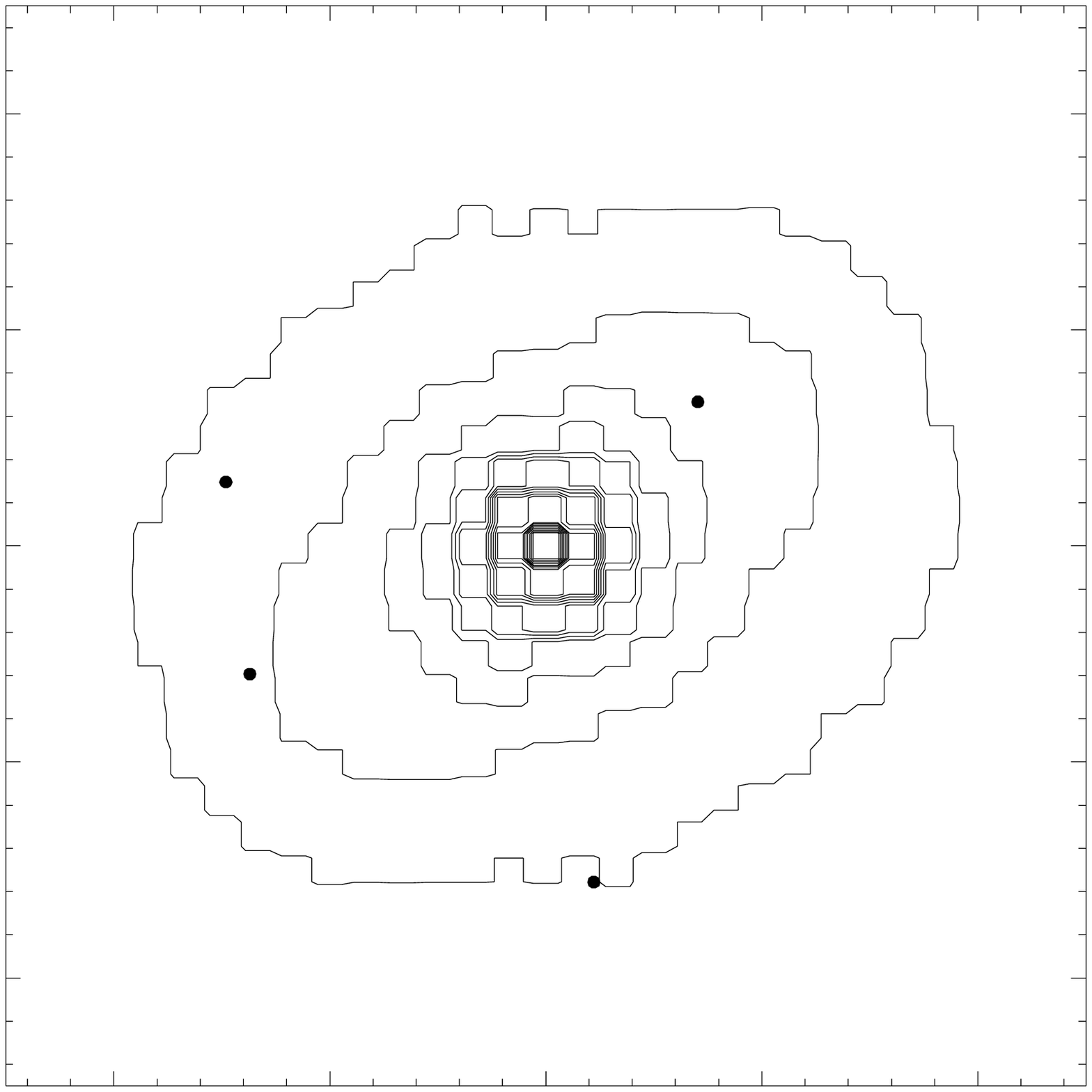}
\caption
{(a) The actual mass profile of galaxy \#1. There is external
shear in this case; its source is located to the upper right of the
galaxy, as can be deduced from the configuration of the images with
respect to the center of the main lensing galaxy. (b) An average of
100 reconstructed mass profiles. In both the panels the contours are 
at $\onethird$, $\twothirds$, 1,\dots, in units of critical density 
for lensing. The four solid dots are the locations of the images.
\label{mass19}}
\end{figure}

\begin{figure}
\plottwo{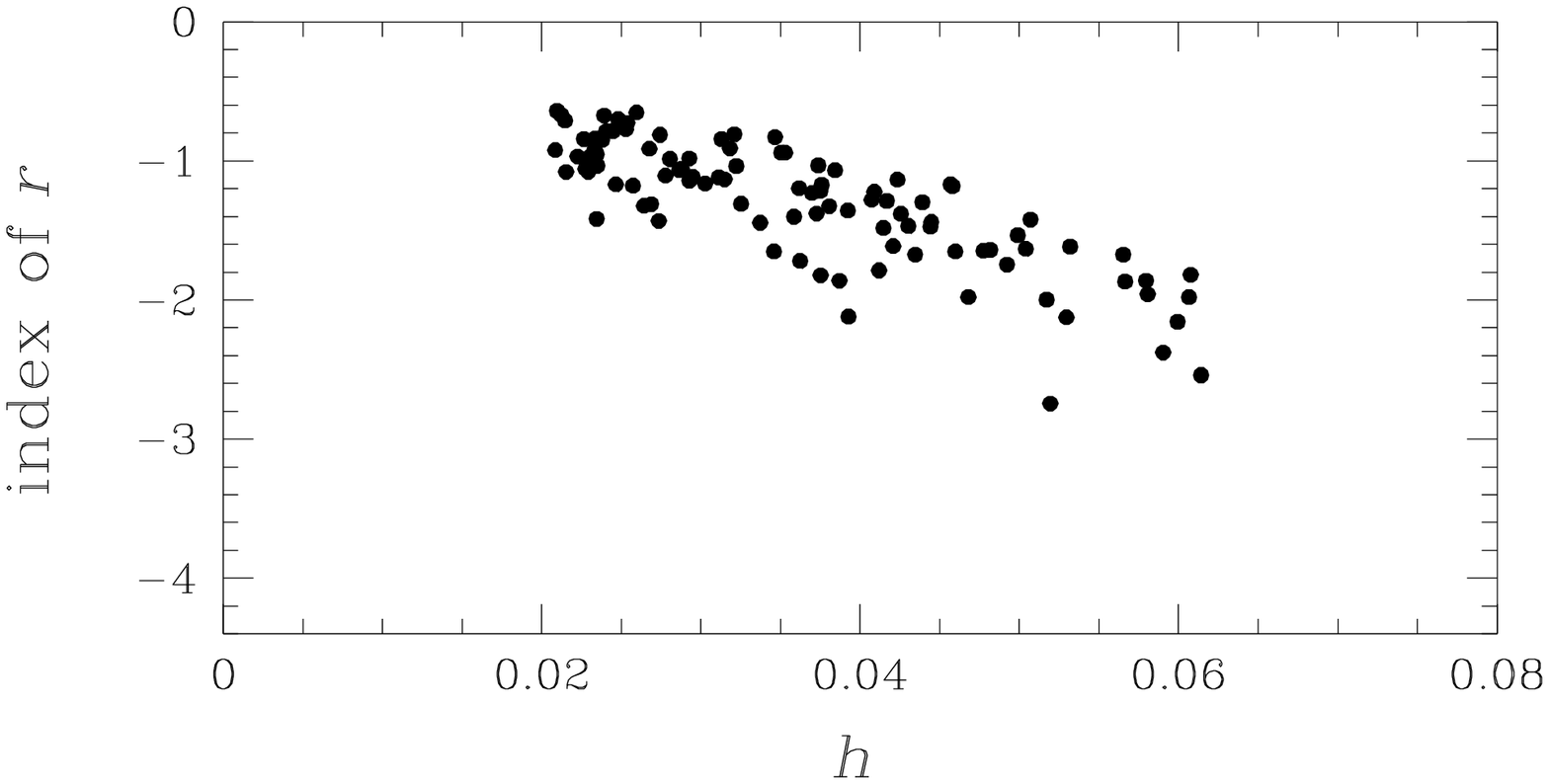}{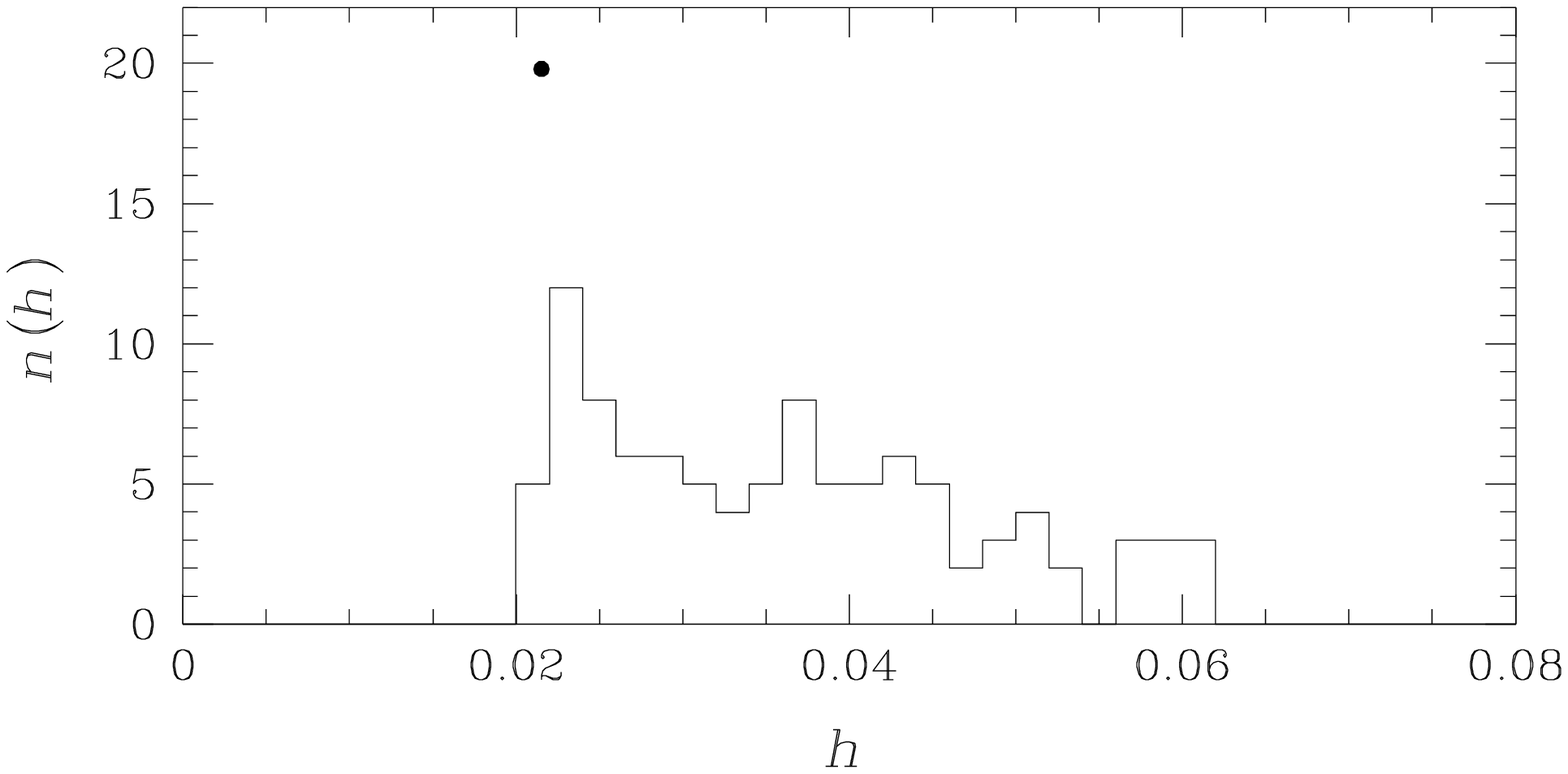}
\caption
{(a) Slope of the projected density profile, 
$d\ln\kappa/d\ln\theta={\rm ind}(r)$ plotted against the derived $h$ value.
(b) Probability distribution of derived $h$ values from galaxy \#1. The
actual value of $h$ is 0.025 for all four blind-test galaxies. The dot
indicates the location of the `most isothermal' reconstructed galaxy. Note
that the whole estimated range spans a factor of 3 in $h$, and the distribution
is not Gaussian.
\label{h19}}
\end{figure}

\begin{figure}
\plottwo{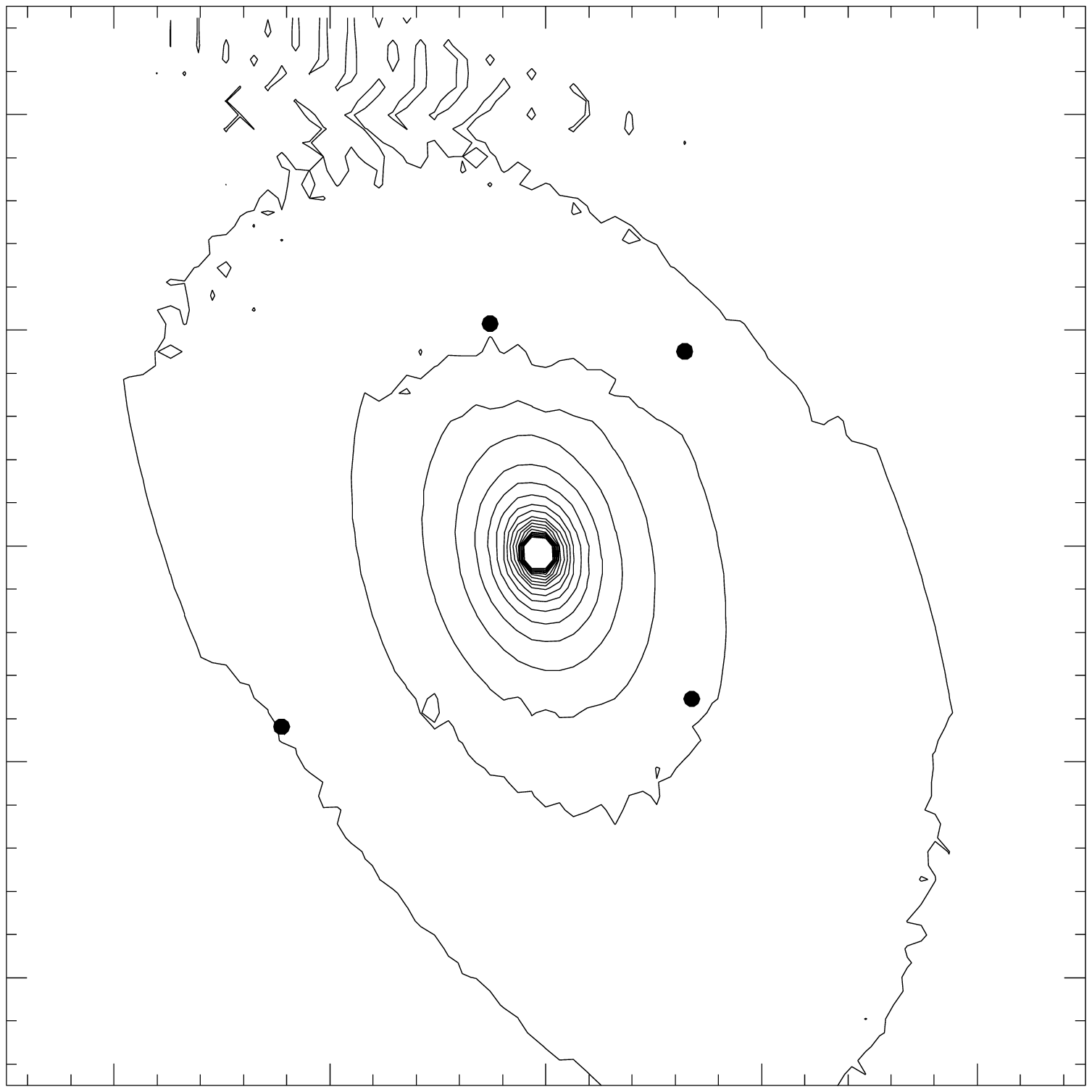}{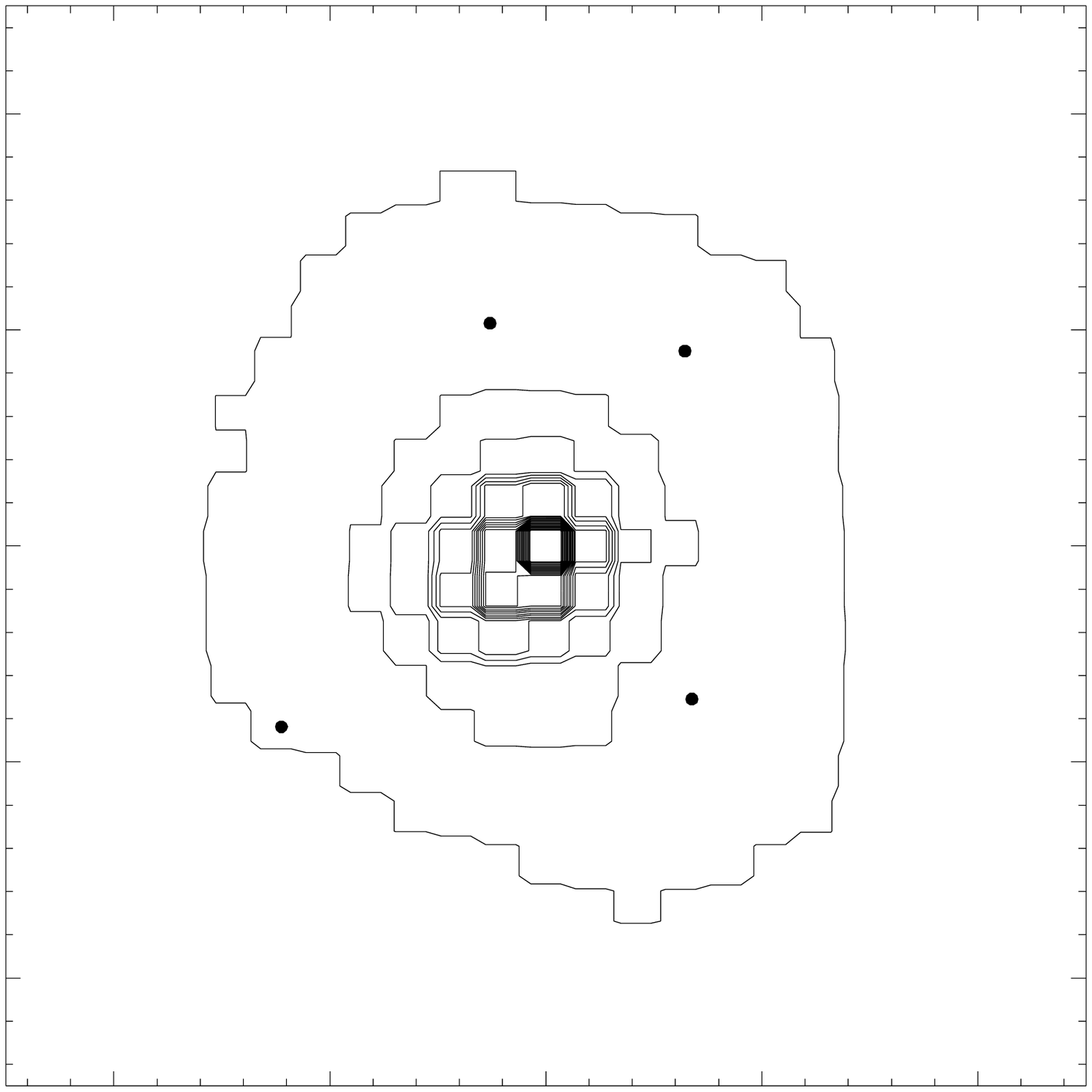}
\caption
{Same as Figure~\ref{mass19} but for galaxy \#2. The source of external
shear is located to the upper right of the main galaxy.
\label{mass20}}
\end{figure}

\begin{figure}
\plottwo{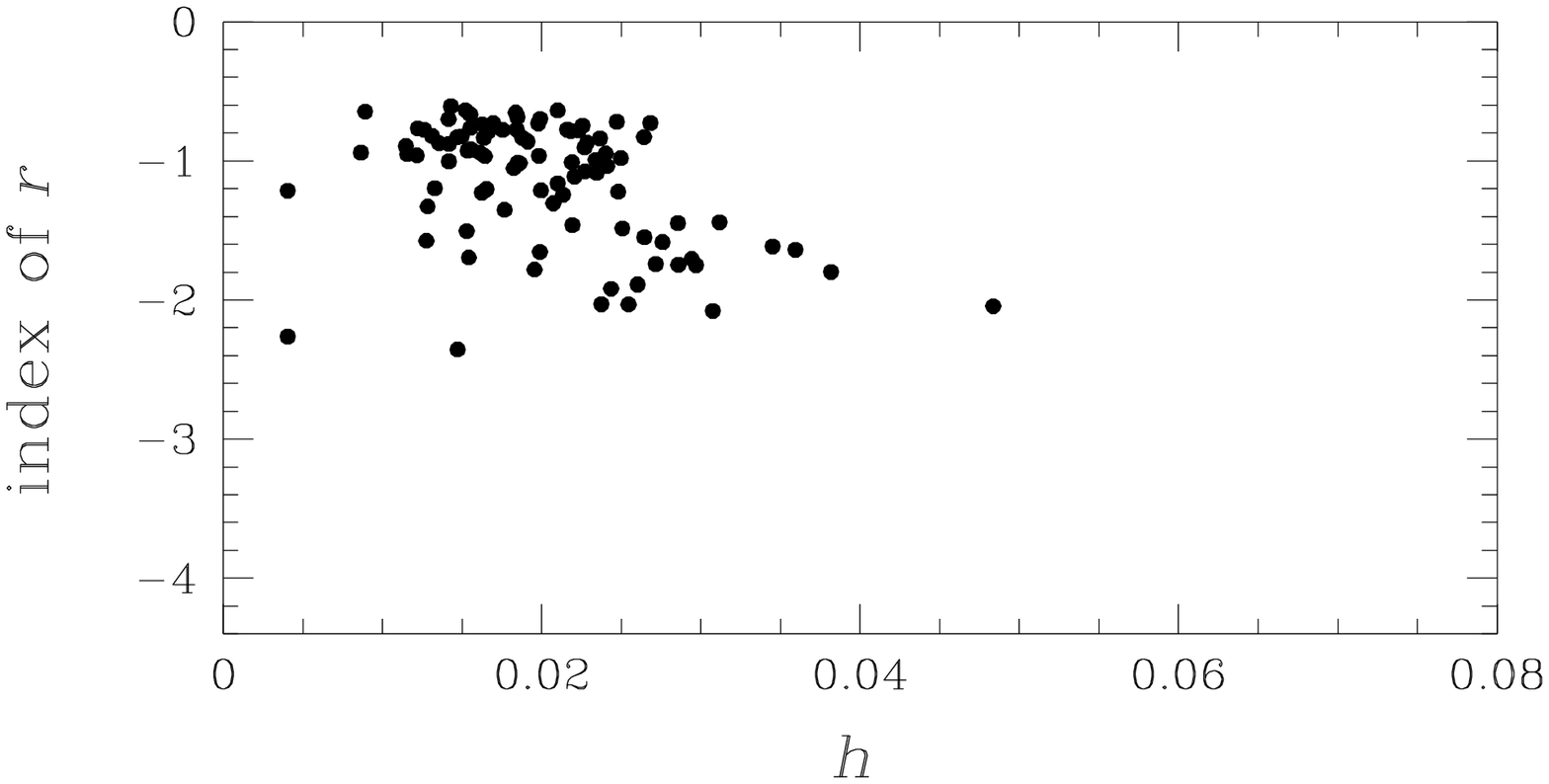}{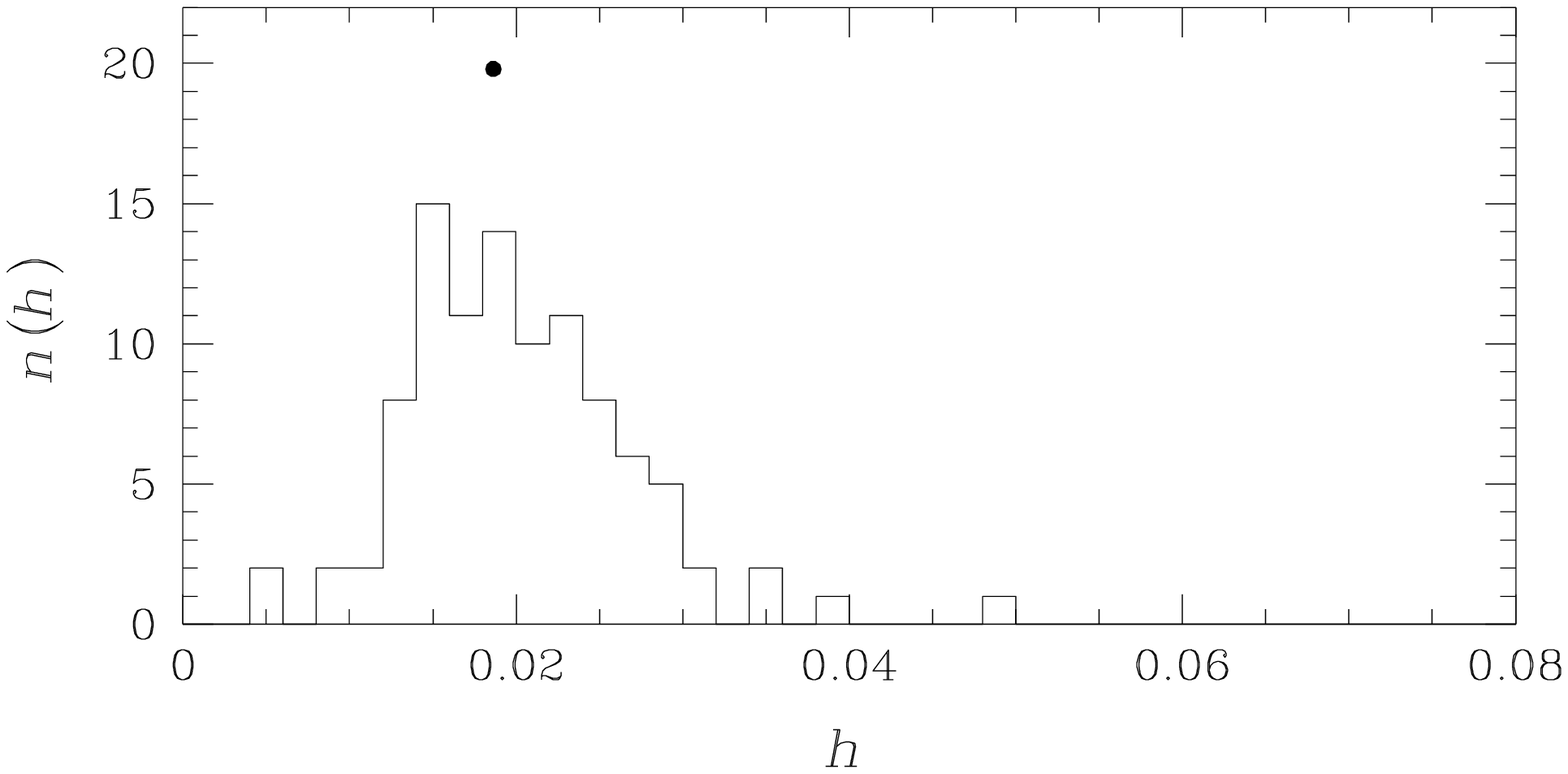}
\caption
{Same as Figure~\ref{h19}, but for galaxy \#2.
\label{h20}}
\end{figure}

\begin{figure}
\plottwo{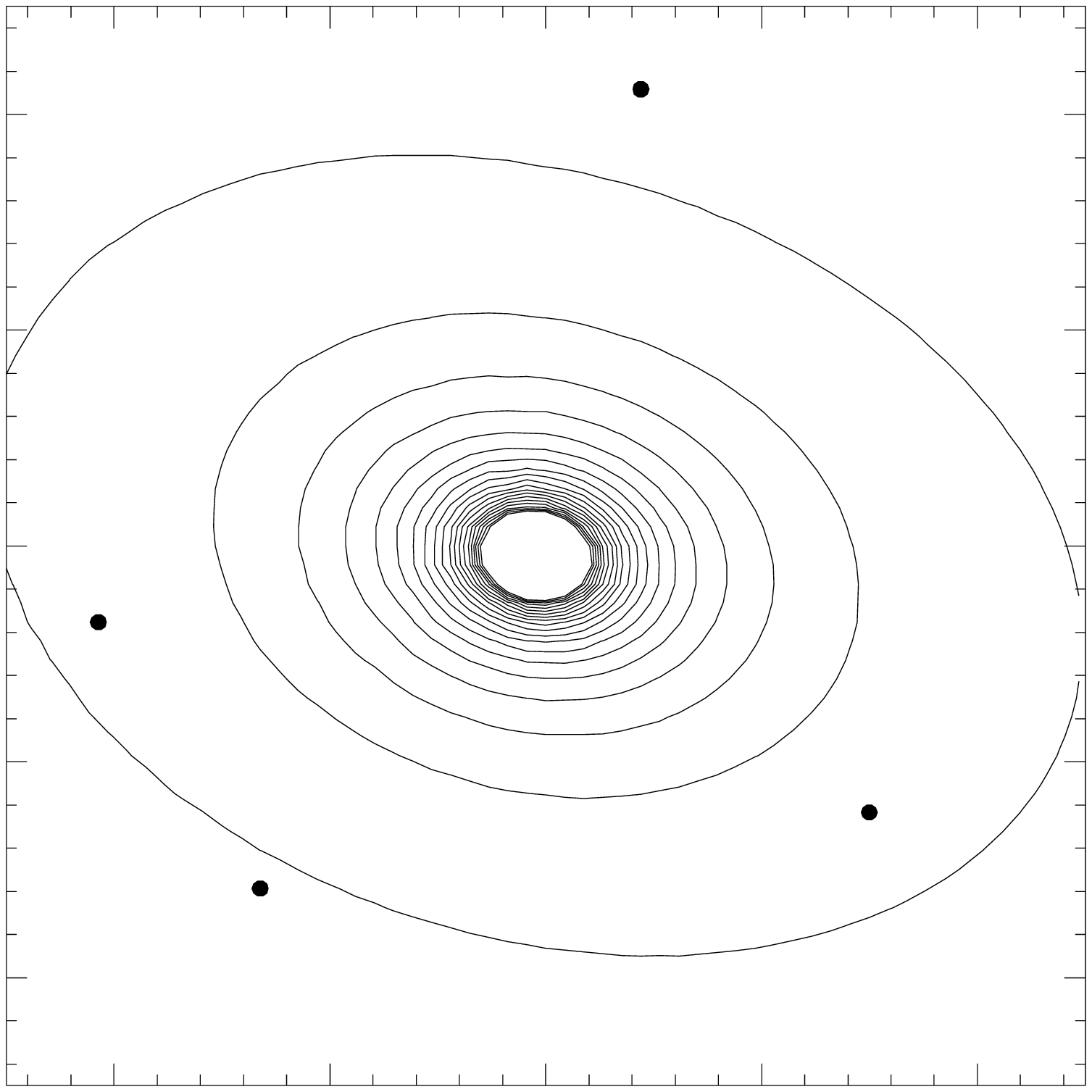}{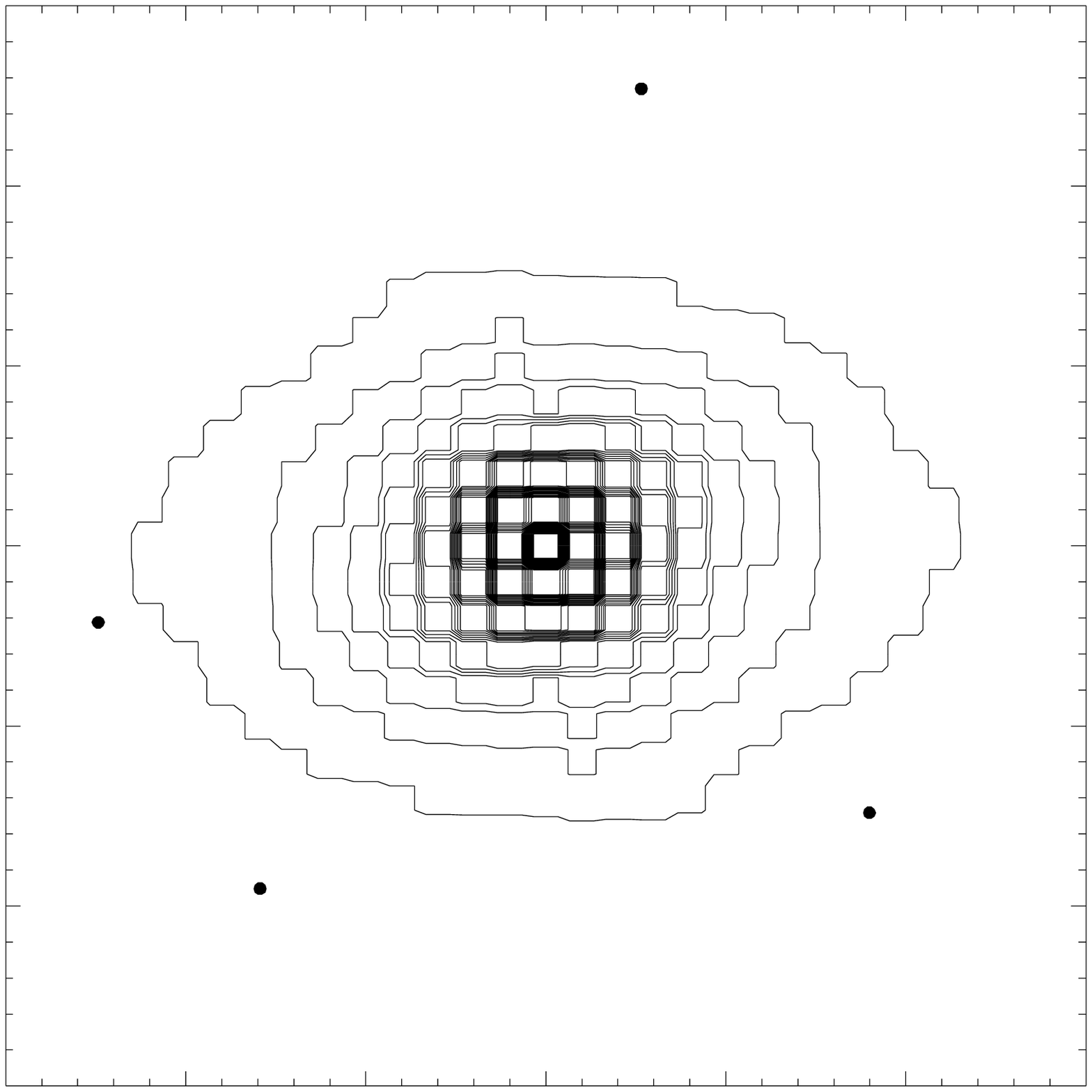}
\caption
{Same as Figure~\ref{mass19} but for galaxy \#3. There is no external
shear in this case.
\label{mass21}}
\end{figure}

\begin{figure}
\plottwo{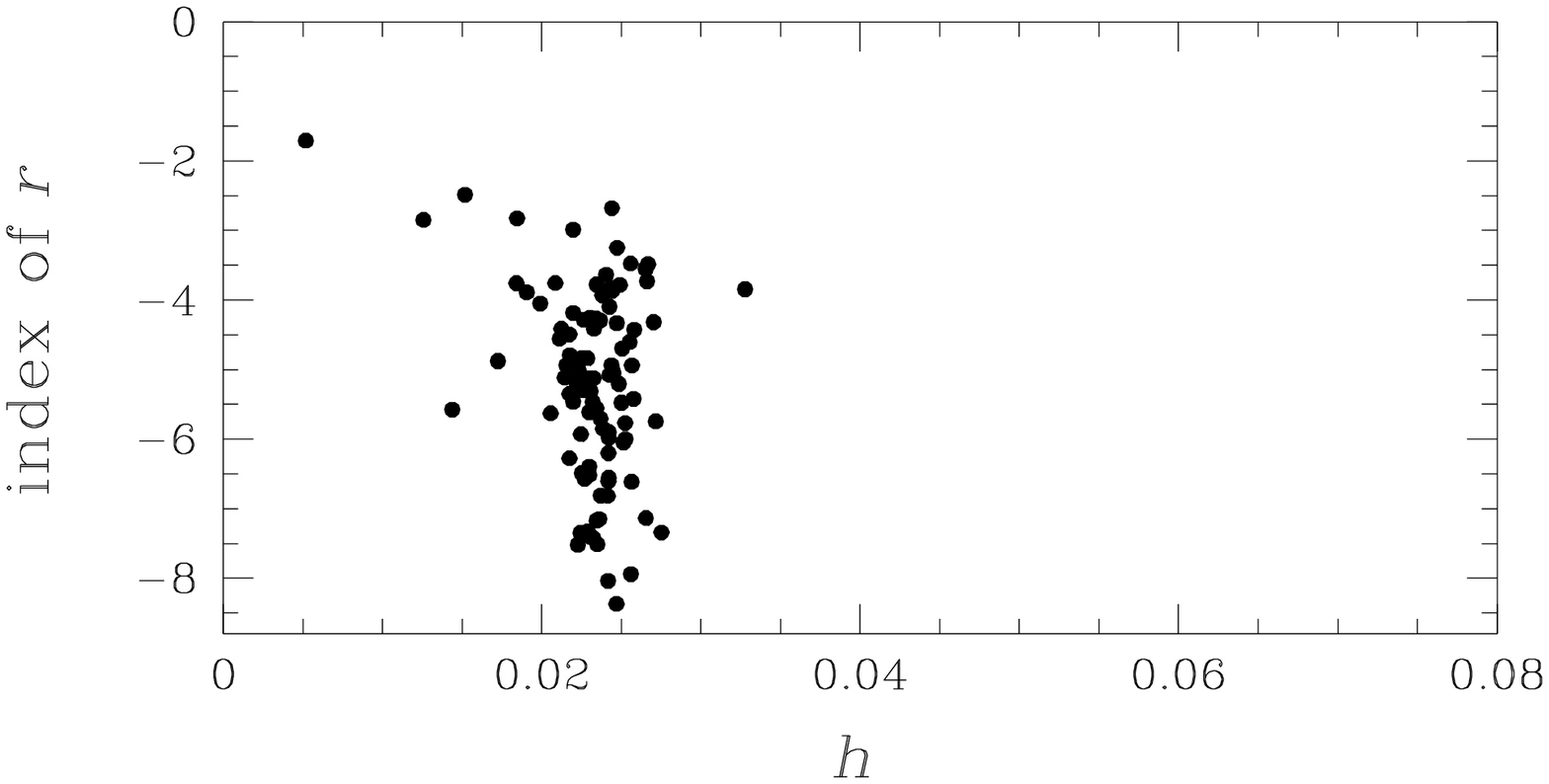}{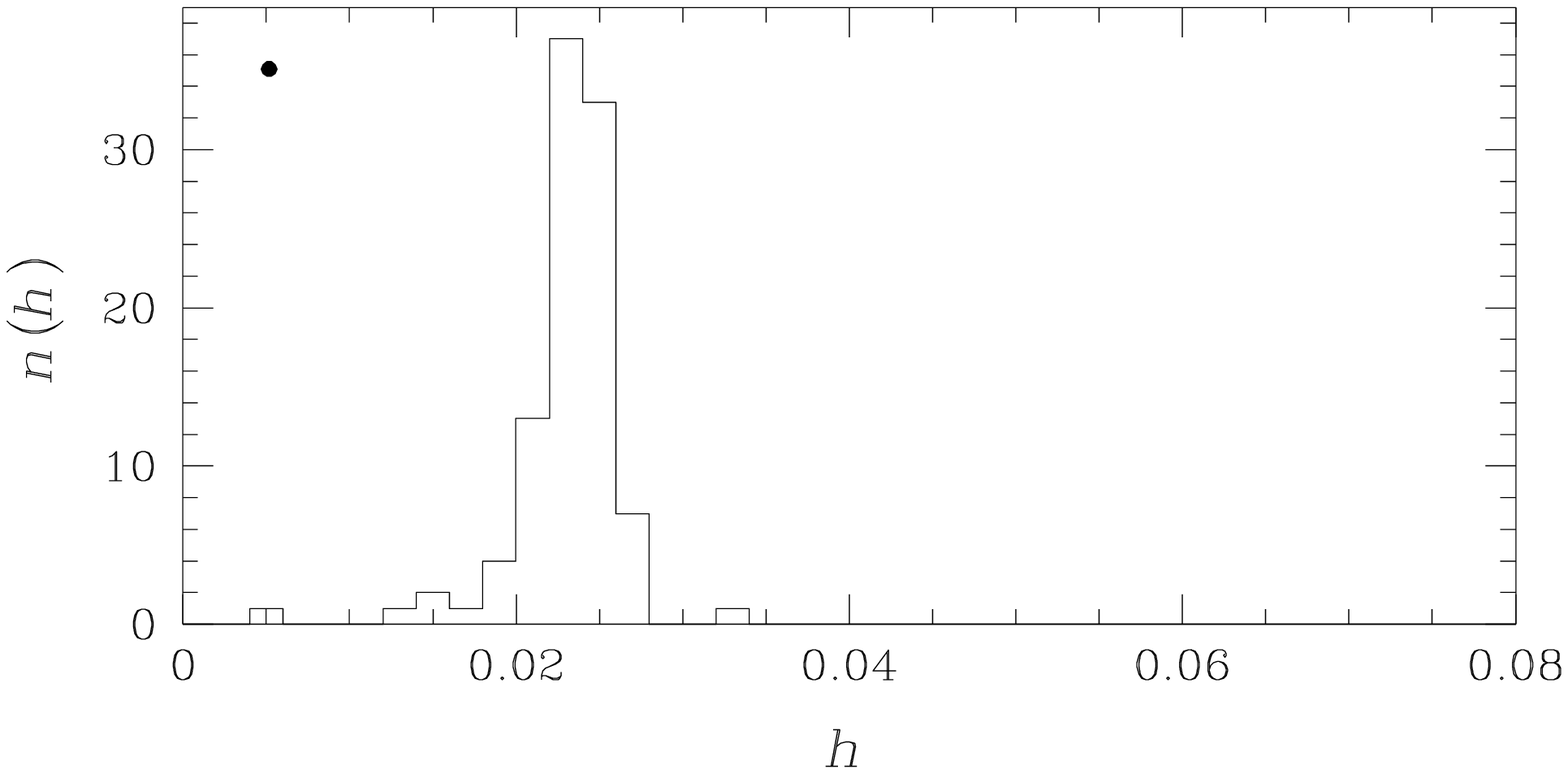}
\caption
{Same as Figure~\ref{h19}, but for galaxy \#3.
\label{h21}}
\end{figure}

\begin{figure}
\plottwo{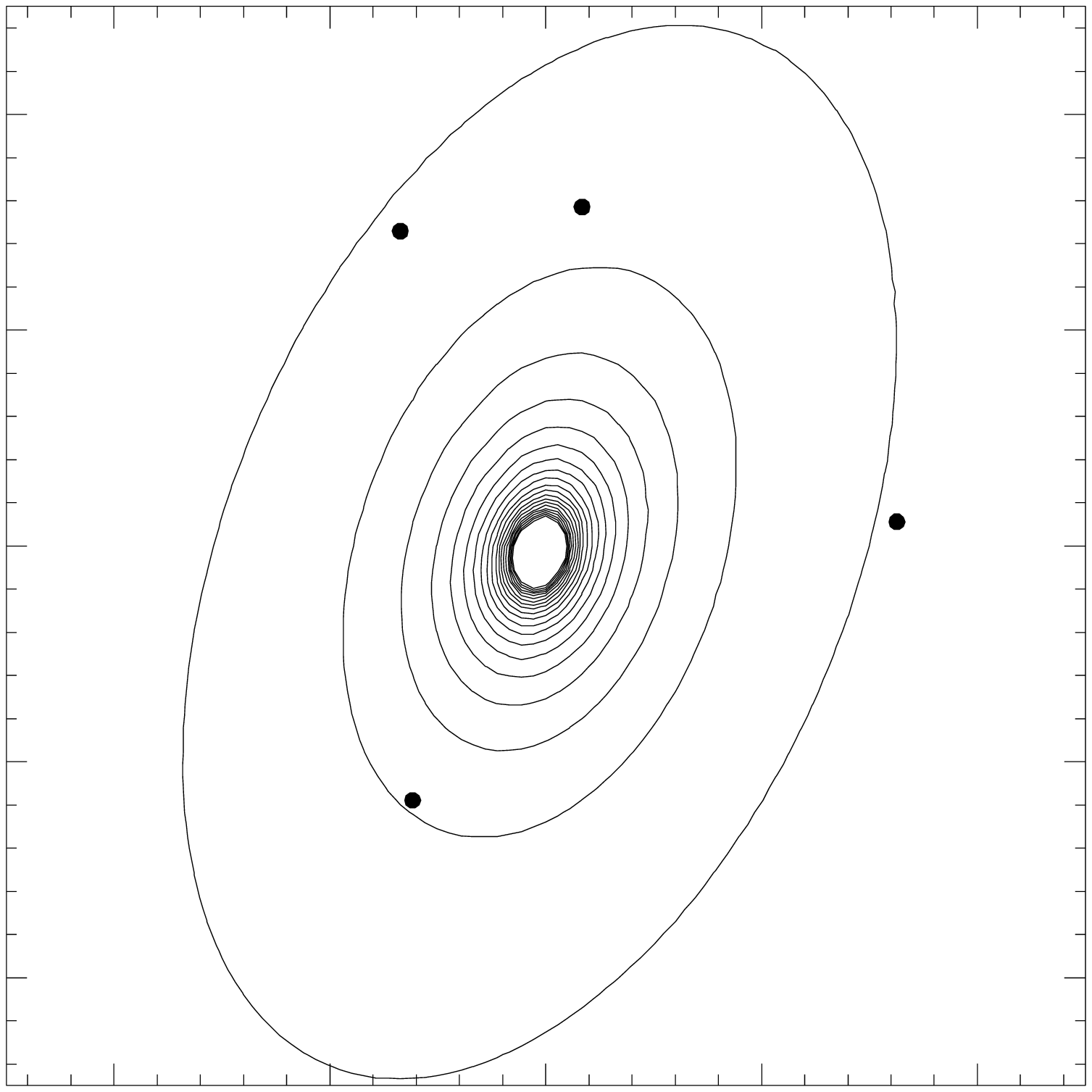}{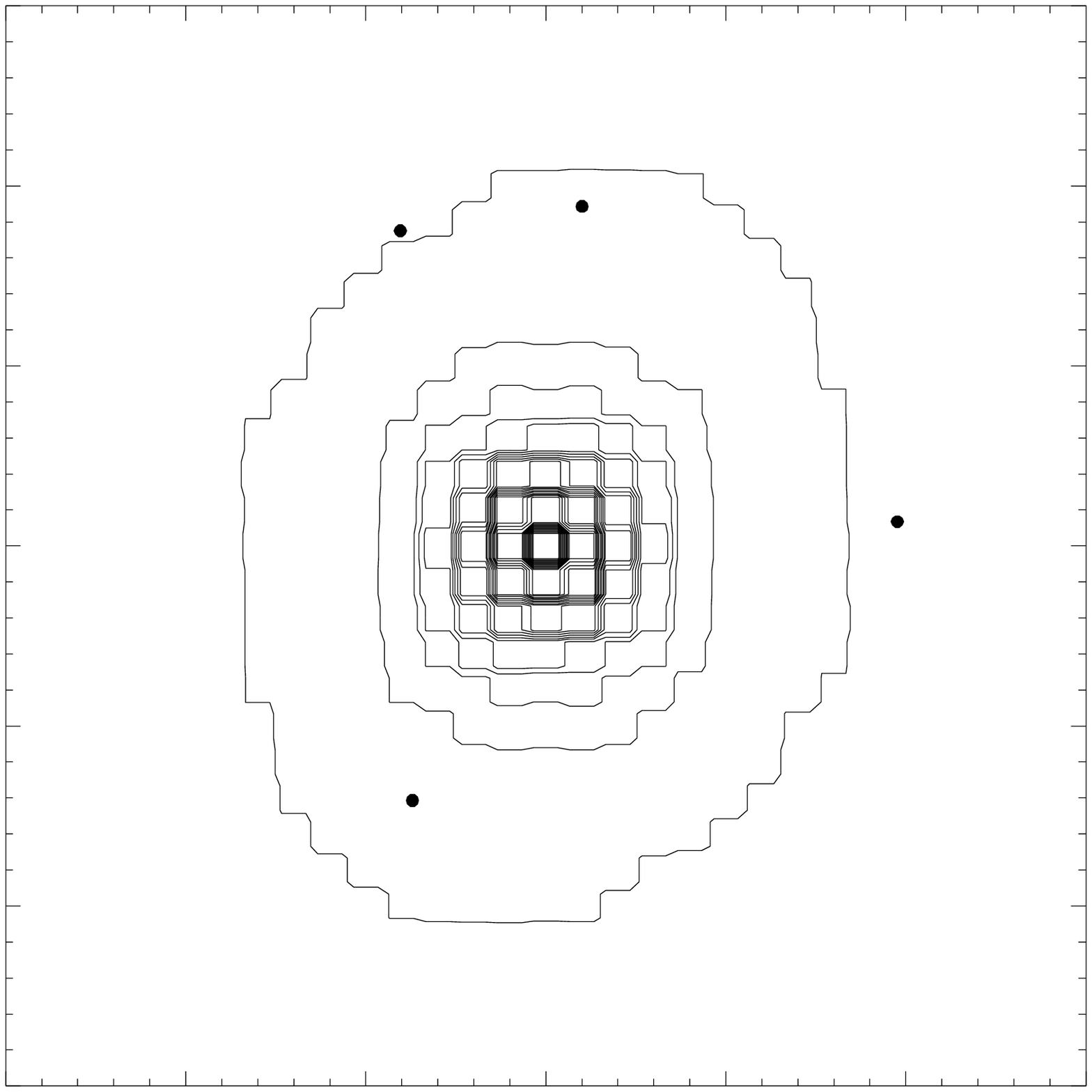}
\caption
{Same as Figure~\ref{mass19} but for galaxy \#4. The source of external
shear is located to the lower left of the main galaxy.
\label{mass22}}
\end{figure}

\begin{figure}
\plottwo{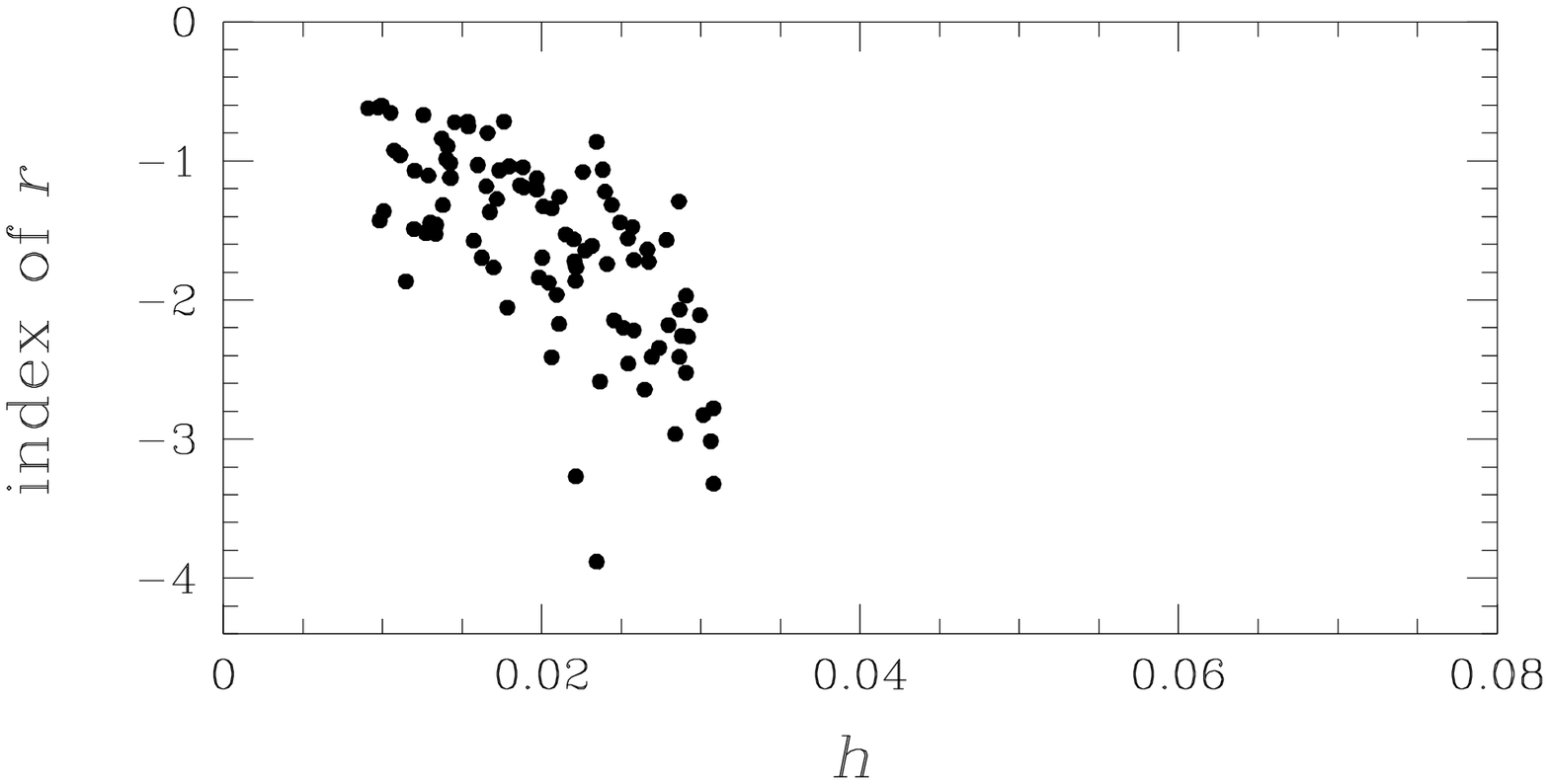}{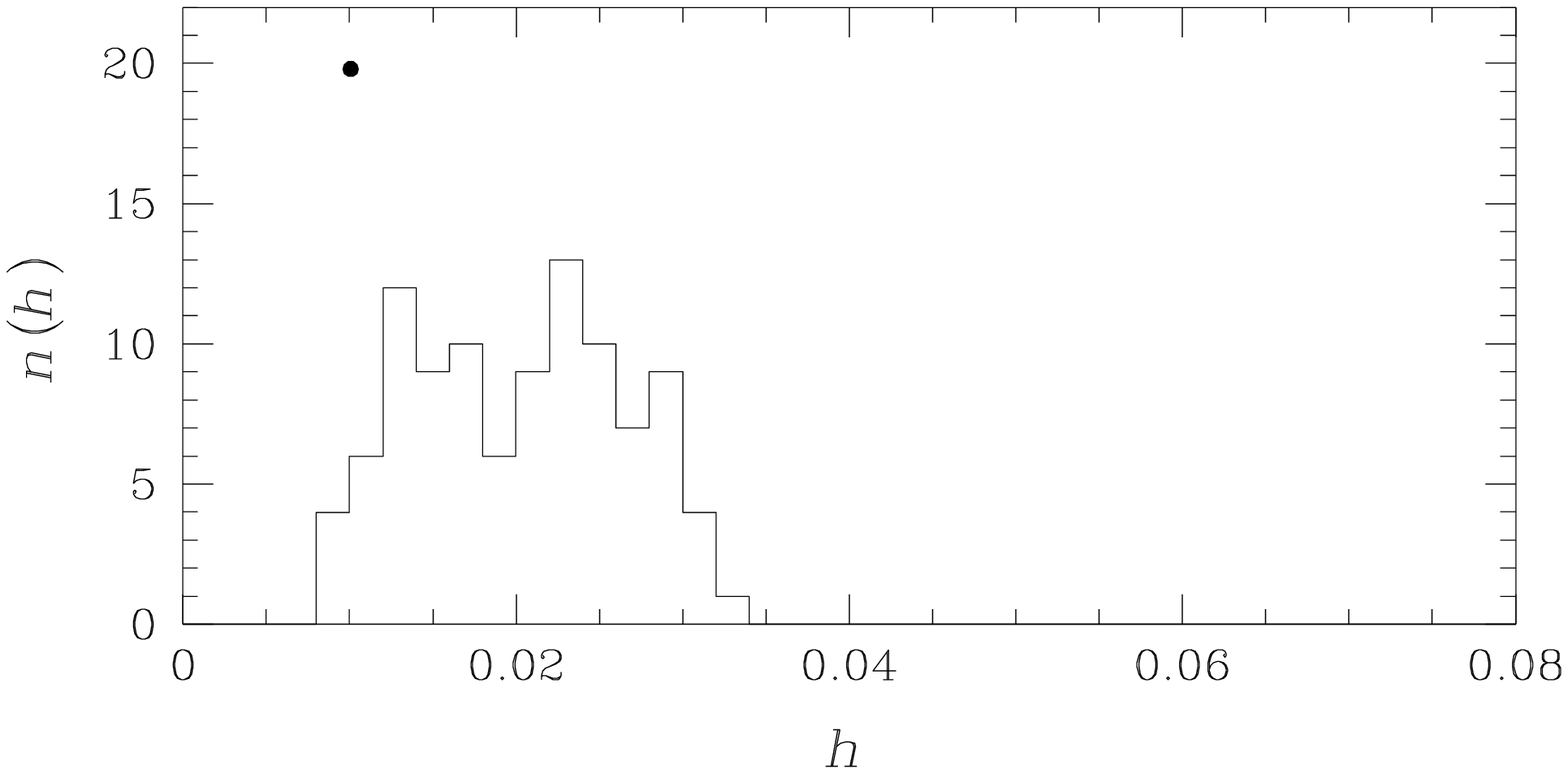}
\caption
{Same as Figure~\ref{h19}, but for galaxy \#4.
\label{h22}}
\end{figure}

Person B applied the reconstruction method to each system twice, once
with the assumption of inversion symmetry (i.e., symmetric galaxies,
see item 4 in Section~\ref{method}), and once without. Based on the
appearance of the reconstructed mass distribution Person B decided
whether inversion symmetry constraint was right in each
case. Figures~\ref{mass19}--\ref{h22} present the results for each of
the four galaxies. For galaxies \#1, 3 and 4 Person B picked symmetric
options, and the asymmetric option for galaxy \#2. Panels (a) and (b)
of Figures~\ref{mass19}, \ref{mass20}, \ref{mass21}, and \ref{mass22}
show the actual projected density distribution and the average of the
100 reconstructed galaxies, for galaxies \#1, 2, 3, and 4
respectively.  In a map which is an {\it average} of many
reconstructions, persistent features of individual maps are enhanced
while peculiarities are washed out, so the average is a reasonable
guess as to what the real galaxy looks like, in a probabilistic sense.

Panels (a) of Figures~\ref{h19}, \ref{h20}, \ref{h21}, and \ref{h22} 
plot the slope of density profile, ${\rm ind}(r)$ vs. derived $h$.
The `real' value of $h$ is 0.025.
In all the cases the slope of the density profile, ${\rm ind}(r)$ in the 
vicinity of the images correlates with the derived $h$ value, though the 
degree of correlation and its slope is not universal. Qualitatively, the 
reason for the correlation is easily understood. A relatively flat galaxy 
density profile, i.e., $|{\rm ind}(r)|$ is small, translates 
into a flat gravitational contribution to the arrival time surface, and 
`fills' the well of the geometrical time delay contribution evenly resulting 
in small fluctuations in the amplitude of the total arrival time surface.
Thus the predicted time delays between images will be small, and to keep the
observed time delays fixed the derived $h$ has to be small as well.

\begin{figure}
\plotone{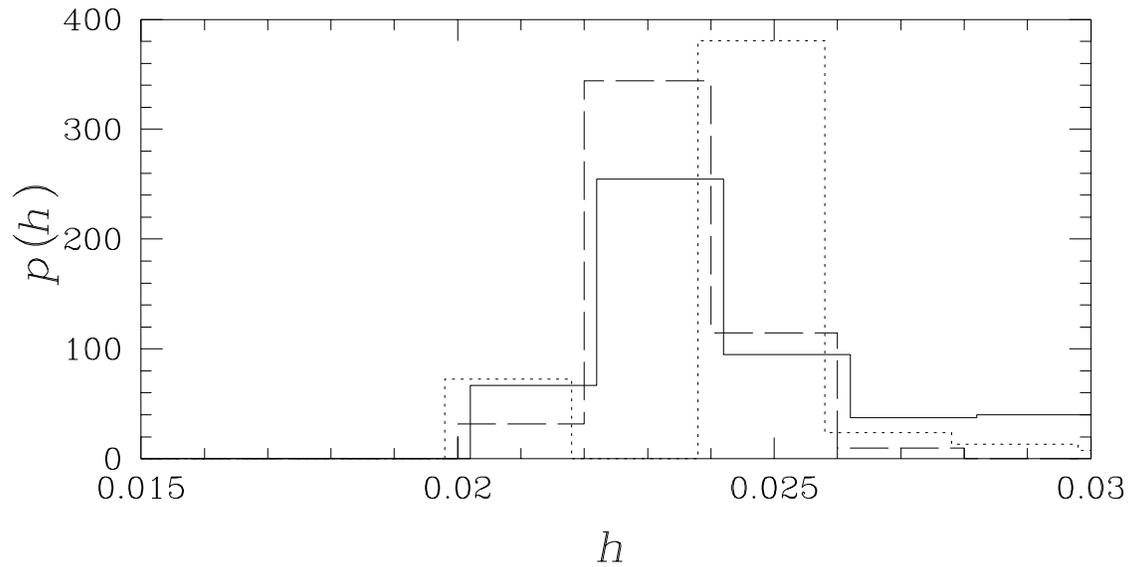}
\caption
{Combined $h$-distributions using independent results from four
blind test galaxies. The solid line represents the case where the modeler
decided on the inclusion/exclusion of the inversion symmetry constraint for
each galaxy separately; the dashed histogram is similar, but excludes the 
best-constrained galaxy (\# 3), and the dotted histogram represents 
the case where inversion symmetry was not applied in any of the systems. The
true value of $h$ is 0.025 for all four galaxies.
\label{phcakes_fig}}
\end{figure}

Panels (b) of Figures~\ref{h19}, \ref{h20}, \ref{h21}, and \ref{h22} 
show the derived $h$ probability distribution. These distributions
look different for all 
galaxies, because galaxy morphologies are different. Since all four are
independent probability distributions based on that galaxy, the overall 
distribution is just the product of the four, see Figure~\ref{phcakes_fig}.
The solid histogram is the product of the four distributions presented
in panels (b) of Figures~\ref{h19}, \ref{h20}, \ref{h21}, and
\ref{h22}.  The dashed histogram is similar, but results from Person B
excluding what appeared to be the best constrained galaxy (\# 3), and
the dotted histogram represents the case where inversion symmetry was
not applied to any of the systems. All three resultant distributions
recover $h$ fairly well, with the $90\%$ of the models contained
within $20\%$ of the true $h$. However the distributions are not the
same; the most probable values are different by $\sim 10\%$.  This
illustrates how a relatively minor feature in modeling constraints,
namely exclusion or inclusion of inversion symmetry, can make a
considerable difference in the estimated $h$ value when the goal is to
achieve precision of $10\%$. Based on this observation we conclude
that assumed galaxy shape in parametric reconstructions plays a major
role in determining the outcome of the $H_0$ determination.

How robust are the results to the changes in other modeling assumptions? 
Changing pixel size by a factor of $\sim 1.5$, and relaxing mass gradient
angle constraint (item 3 in Section~\ref{method}) does not change our
results considerably.

\section{Application to real systems}\label{real}

\subsection{PG1115}\label{PG1115}
\begin{figure}
\plotfour{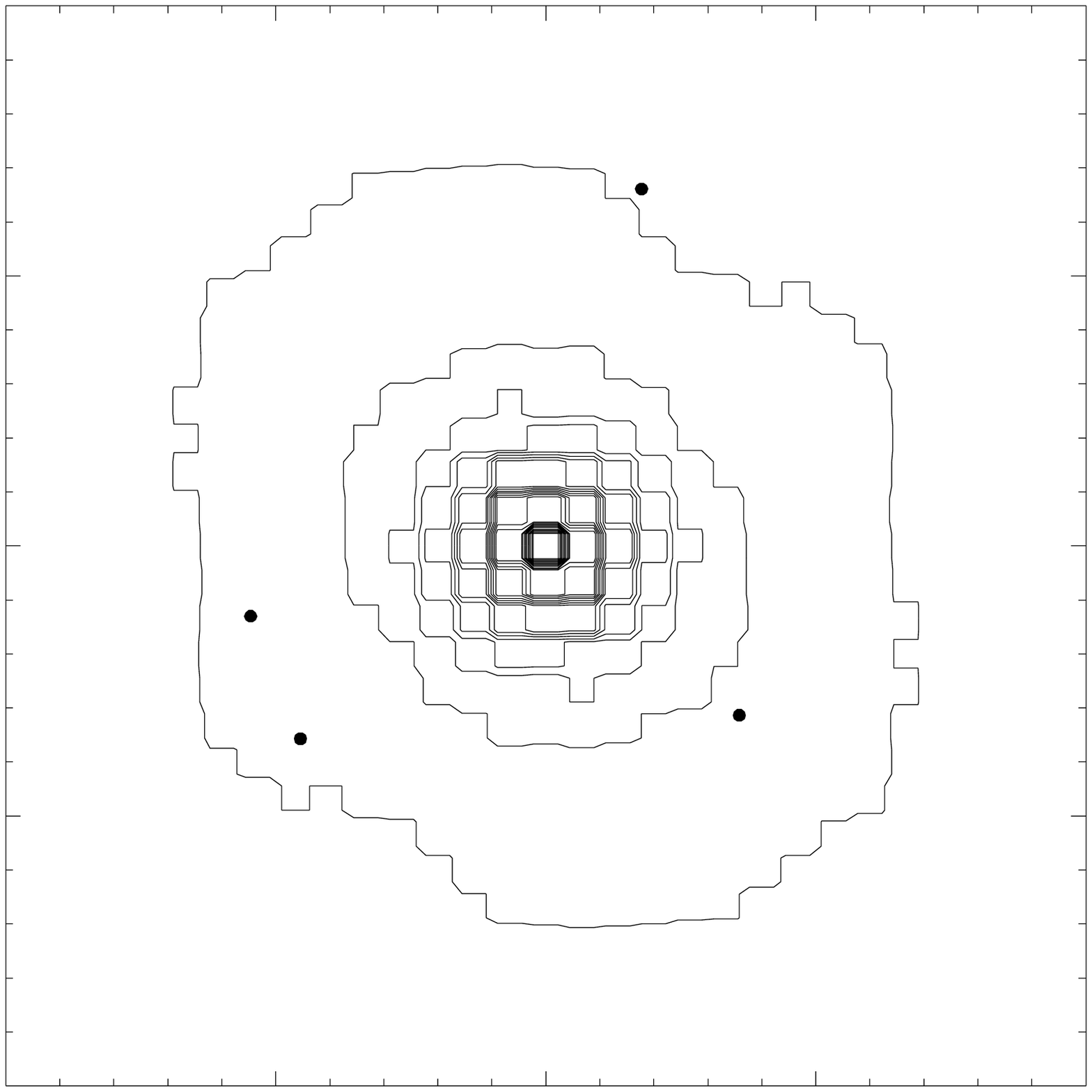}{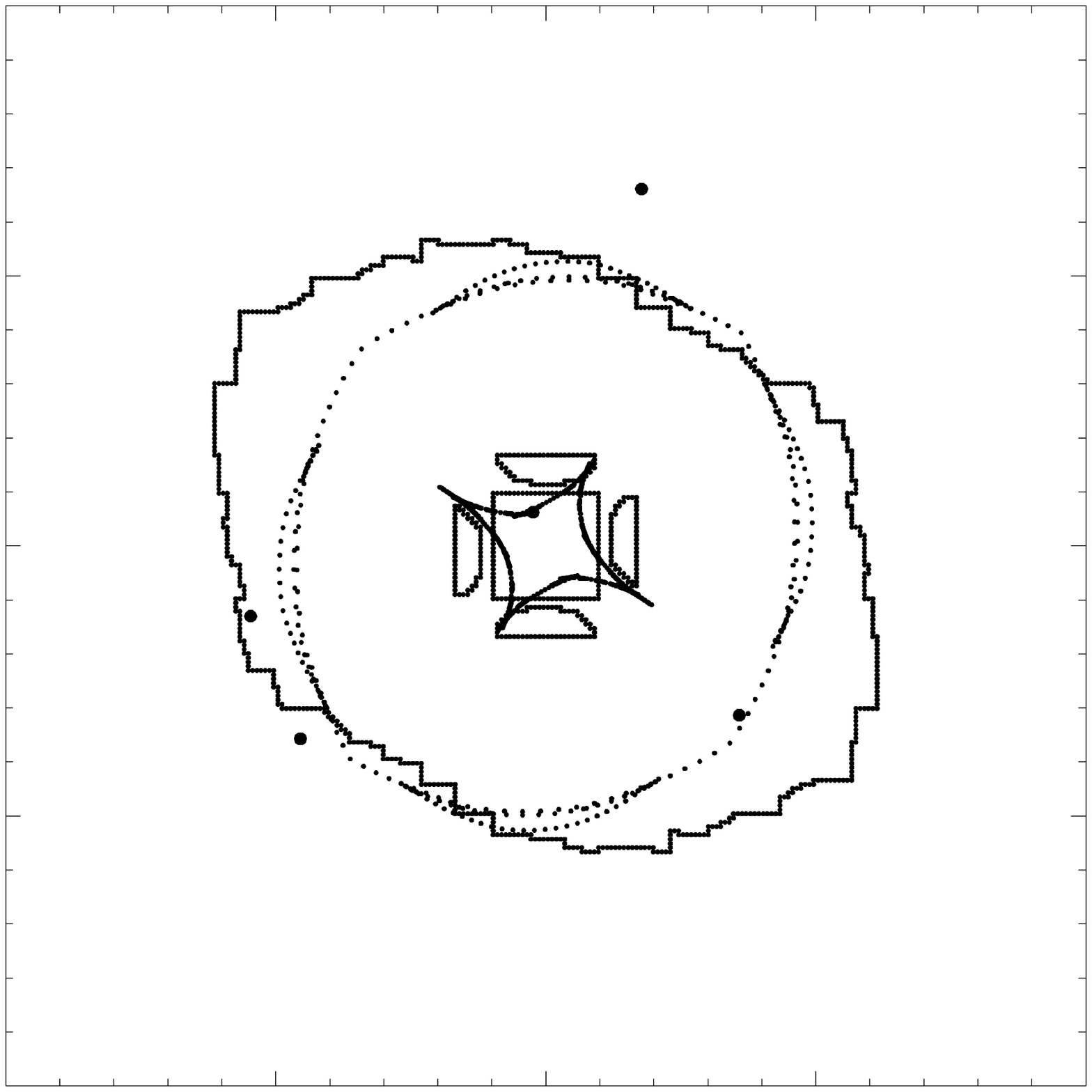}{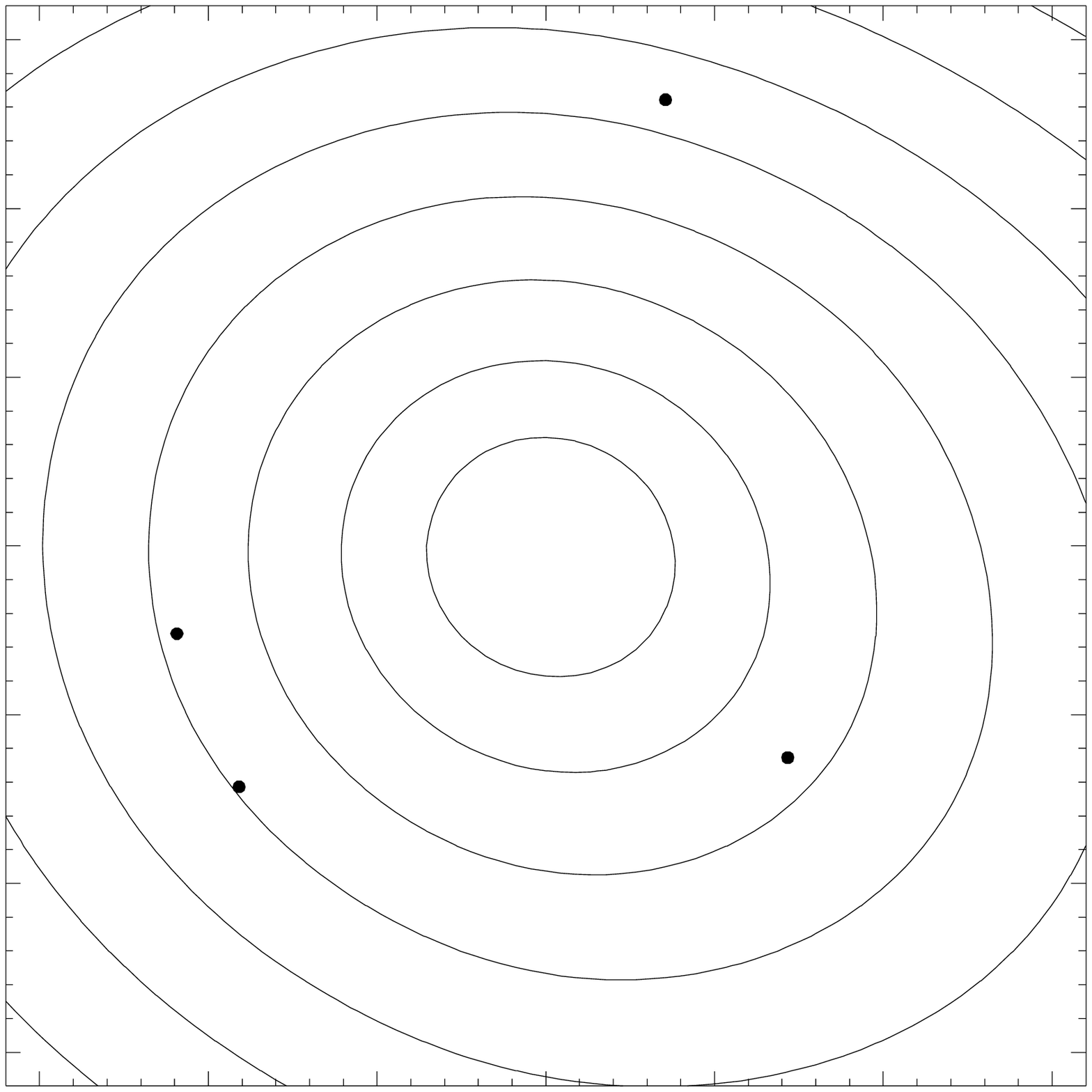}{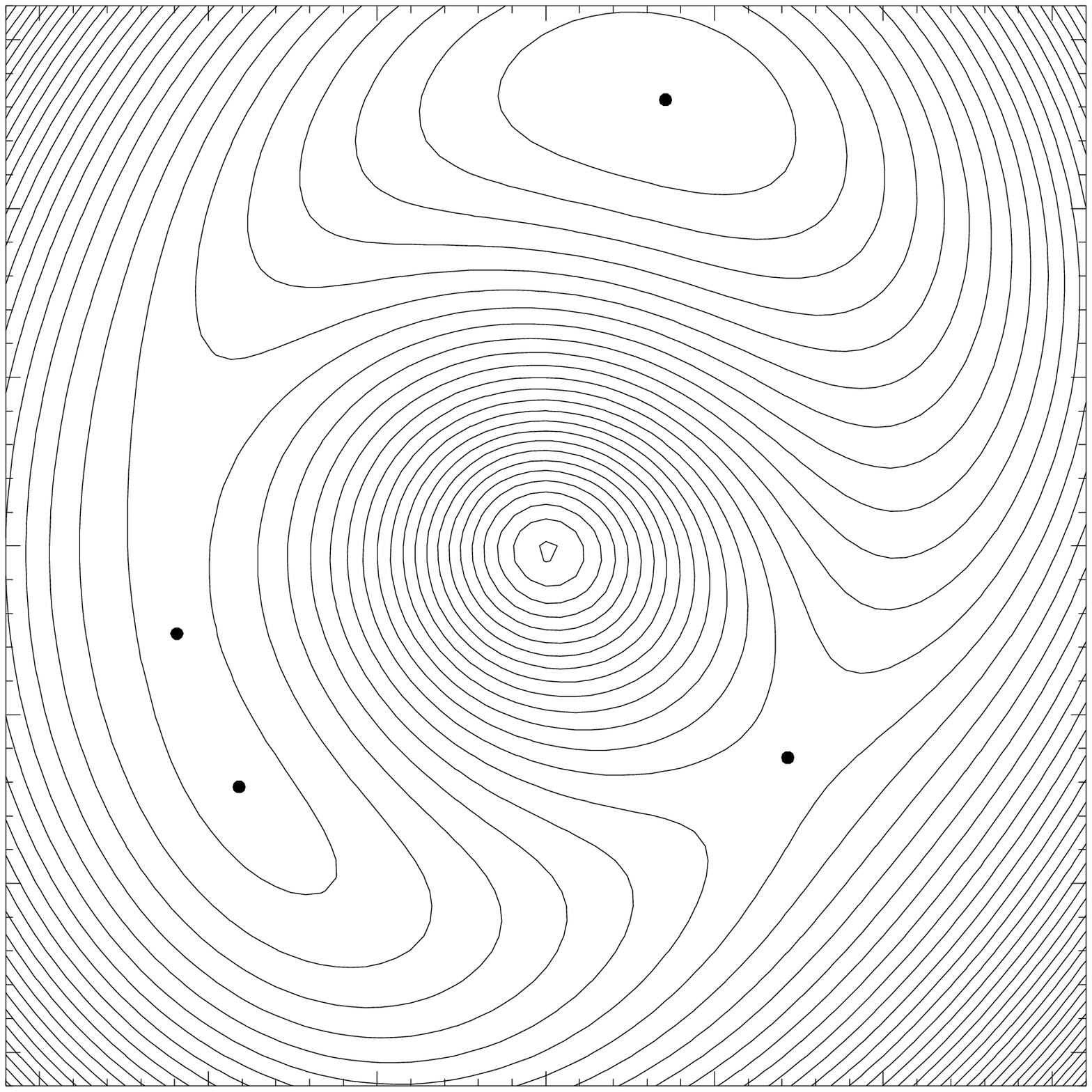}
\caption
{Reconstruction of the galaxy in PG1115+080. Each panel shows averages of 
100 reconstructions; image positions are represented by solid dots. 
(a) Mass map; (b) Caustics (smooth dotted lines) and critical lines (jagged); 
(c) Lensing potential; (d) Arrival time surface.
\label{four_1115}}
\end{figure}

Figure~\ref{four_1115}
shows the results of the reconstruction for PG1115. Since
the main lensing galaxy has no close companions and its light profile 
is smooth we have included inversion symmetry as one of the modeling
constraints. The average of 100 arrival time surfaces is shown in 
Figure~\ref{four_1115}(a); Figure~\ref{four_1115}(b)
shows the corresponding caustics and critical lines. The latter
are not as smooth as the former because locations of caustics and critical
lines are derived using the gradients the arrival time surface, which are
always noisier than the original function. Panels (c) and (d) plot the
quantity $\vtheta\cdot\vthes-\sum_n \kappa_n \psi_n(\vtheta)$,
and the total arrival time surface, respectively. The plot of the modified
gravitational potential, (c), illustrates the effect of external shear
which is due to a galaxy group to the lower right of the main galaxy. 

\begin{figure}
\plottwo{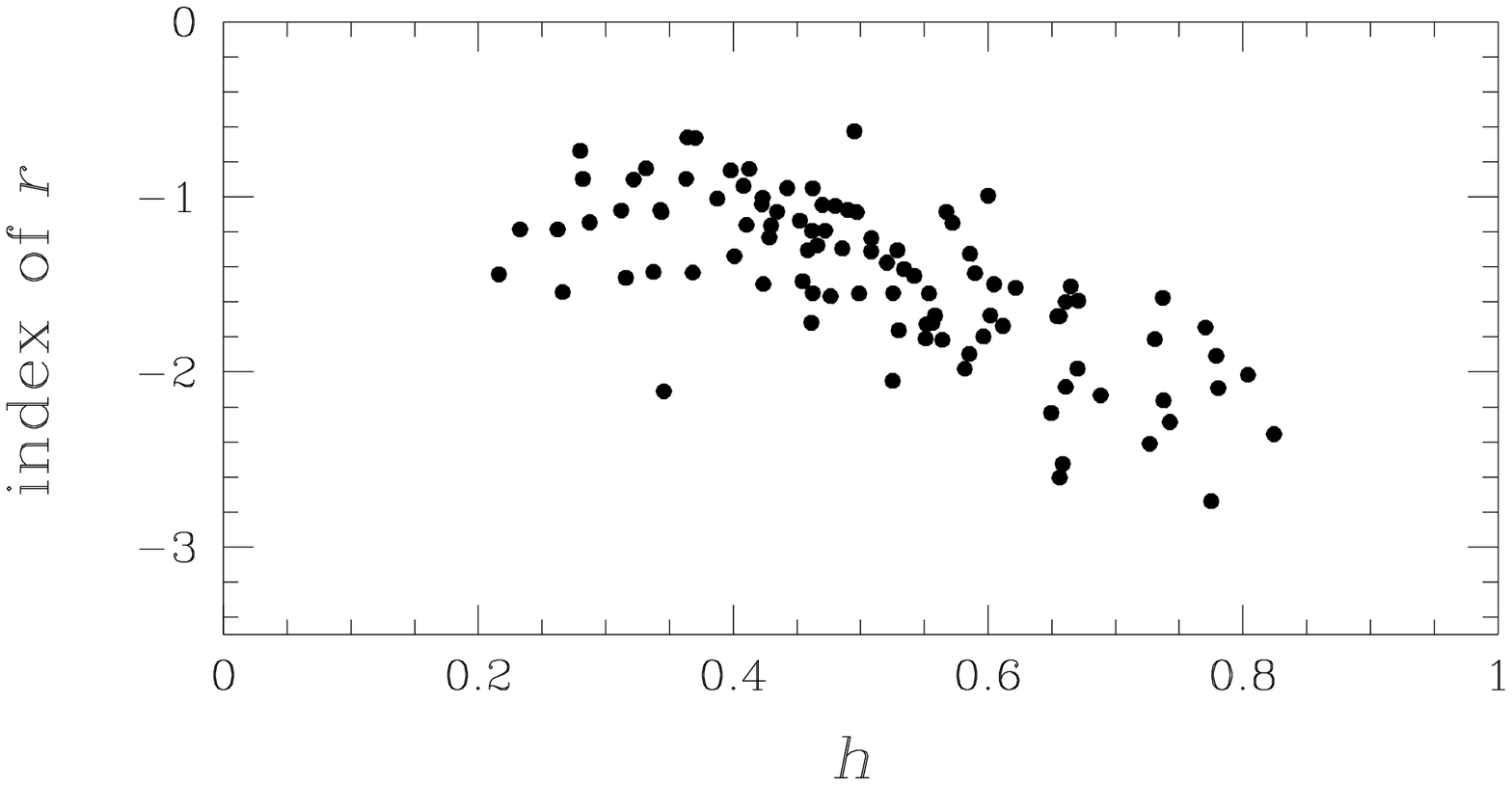}{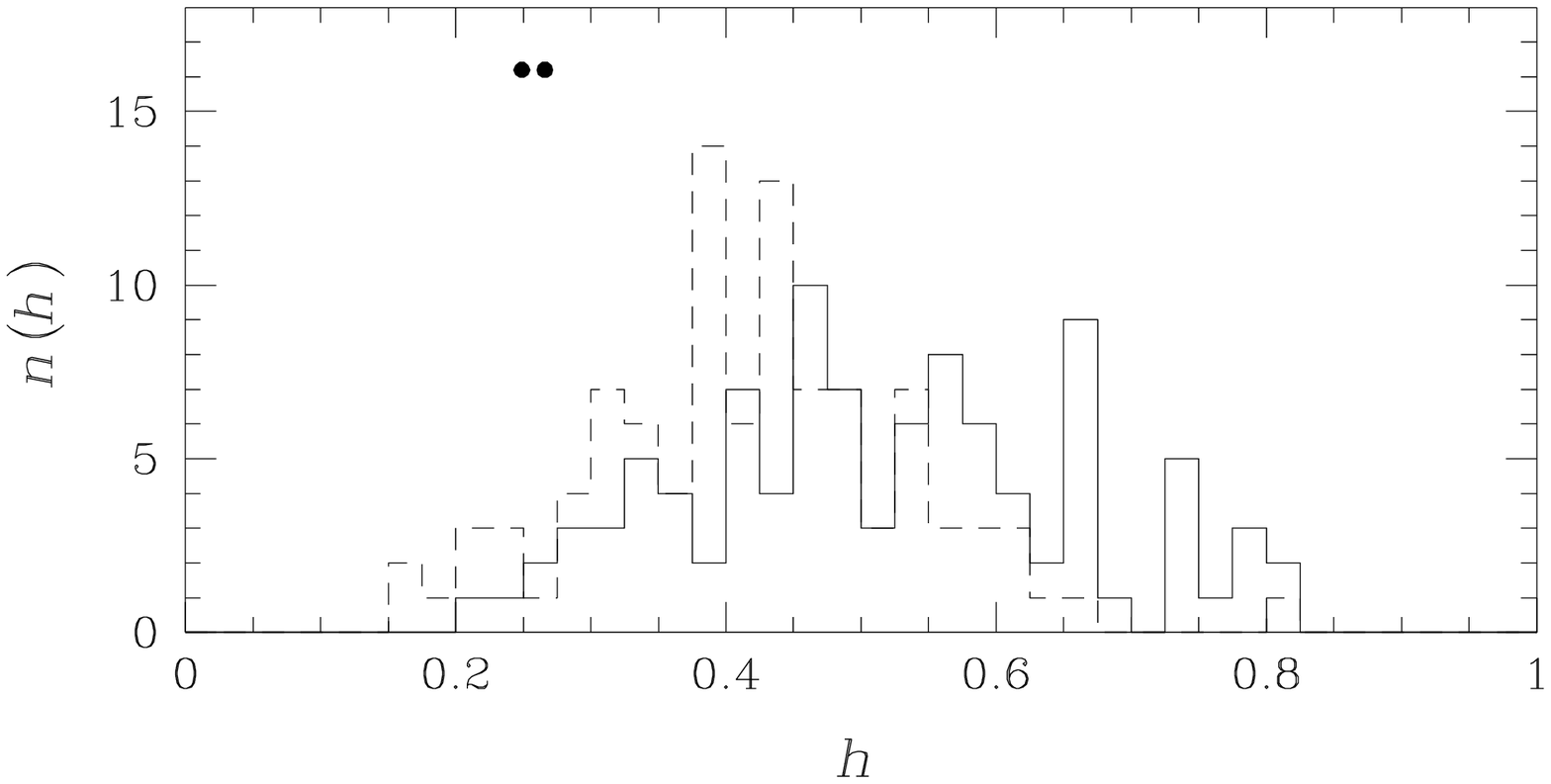}
\caption
{PG1115+080: (a) Slope of the projected density profile,
$d\ln\kappa/d\ln\theta={\rm ind}(r)$ plotted against the derived $h$ value;
(b) Probability distribution of derived $h$ values. Solid and dashed 
histograms use time delays from Barkana (1997), and Schechter et al. (1997)
respectively. The dots mark the location of the `most isothermal' of the
reconstructed galaxies.
Both sets of time delays were derived based on the same
observational data, lightcurves from Schechter et al. (1997).
The difference in the derived distributions illustrates the magnitude of the
error that can arise from errors in time delays measurements.
\label{two_1115}}
\end{figure}

Because ${\rm ind}(r)$ has been measured for the main lensing galaxy in 
PG1115, the relation between profile slope and derived $H_0$, 
Figure~\ref{two_1115}(a), can be used to derive an upper limit on $H_0$. 
Impey et al. (1998) fit the galaxy light 
with a de Vaucouleurs profile of an effective radius $r_e=0.59''$. At the 
location of the images, about $1.3''$ from galaxy center the double 
logarithmic density slope is ${\rm ind}(r)=-2.3$. Assuming that the mass profile
can only be shallower than the light profile, and consulting 
Figure~\ref{two_1115}(a) we place an upper limit on $H_0$ of 
75$\rm\,km\,sec^{-1}\,Mpc^{-1}$. 

If the true mass density profile slope is isothermal the corresponding
$H_0$ is 30$\rm\,km\,sec^{-1}\,Mpc^{-1}$. A low value of $H_0$ was also obtained
by parametric models that assumed isothermal models for the galaxy
(Schechter et al 1997).

In the blind test, Section~\ref{blind}, we assumed that all time delays
are known precisely, which is not currently the case for any of the systems
except Q0957. What effect does an error in time delay determination have on
the derived $H_0$? Figure~\ref{two_1115}(b) shows two distributions derived 
using two different $\Delta t$ determinations based on the same lightcurves.
There is a $20\%$ difference in the most probable value of $H_0$ in
the two histograms, but overall they are not very different. 
Both distributions are very broad; $90\%$ of the models span the range between 
30 and 75$\rm\,km\,sec^{-1}\,Mpc^{-1}$.

\begin{figure}
\plotone{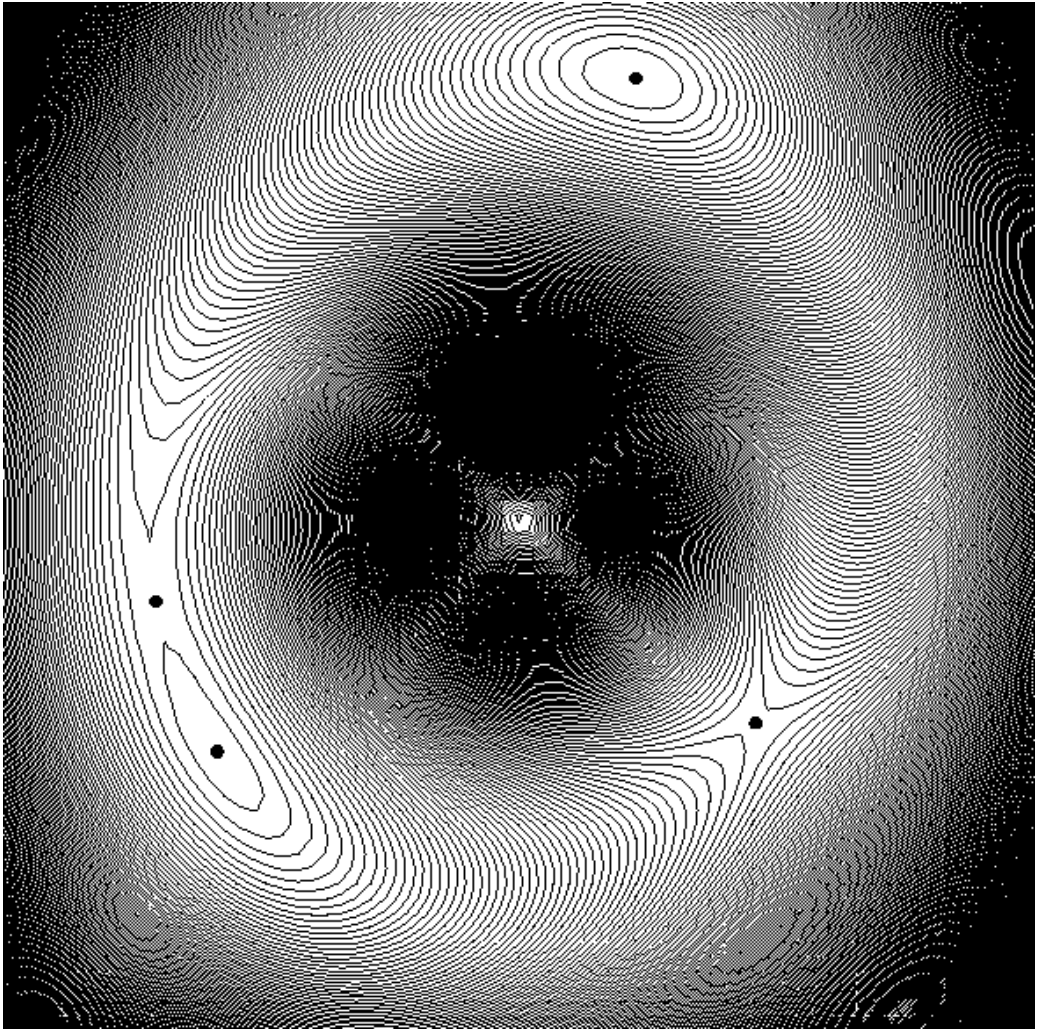}
\caption
{Reconstructed arrival time surface for PG1115 with closely spaced
contours. The flatter regions of the plot outline the optical ring formed
by the four merging images of the faint host galaxy of PG1115 QSO.
\label{ring_1115_fig}}
\end{figure}

Figure~\ref{ring_1115_fig} shows a dense version of the arrival time surface. 
The regions of the plot where the contours are
sparse are the flattest, i.e., most `stationary' regions of the lens plane.
This is where one would expect to find images of sources placed close to the
main source. For example if the point-like QSO is surrounded by a host galaxy
the image of that galaxy will be well delineated by these `empty' regions.
In fact, the observed optical ring in the case of PG1115 is well reproduced
by the ring in Figure~\ref{ring_1115_fig}.

\subsection{B1608}
\begin{figure}
\plotfour{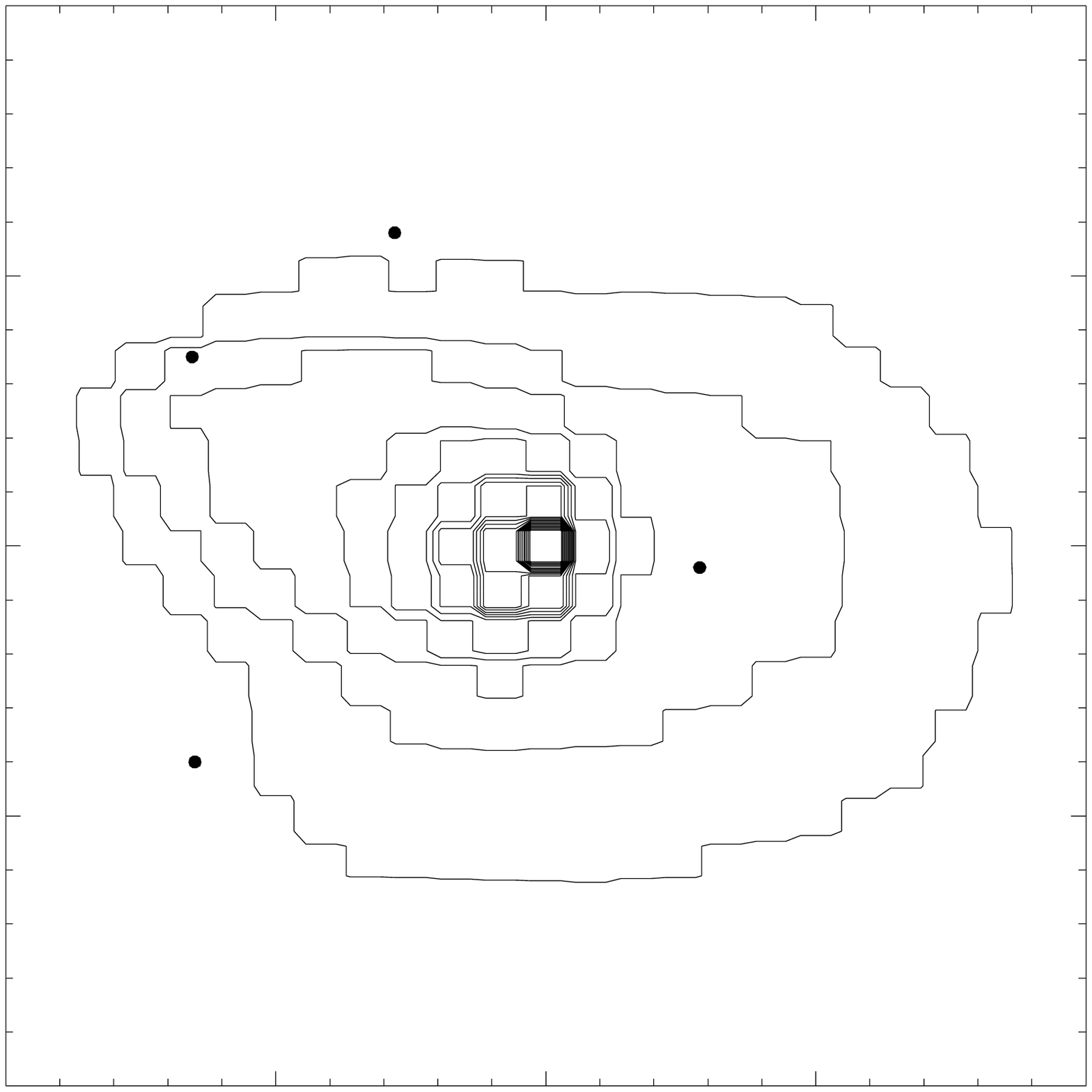}{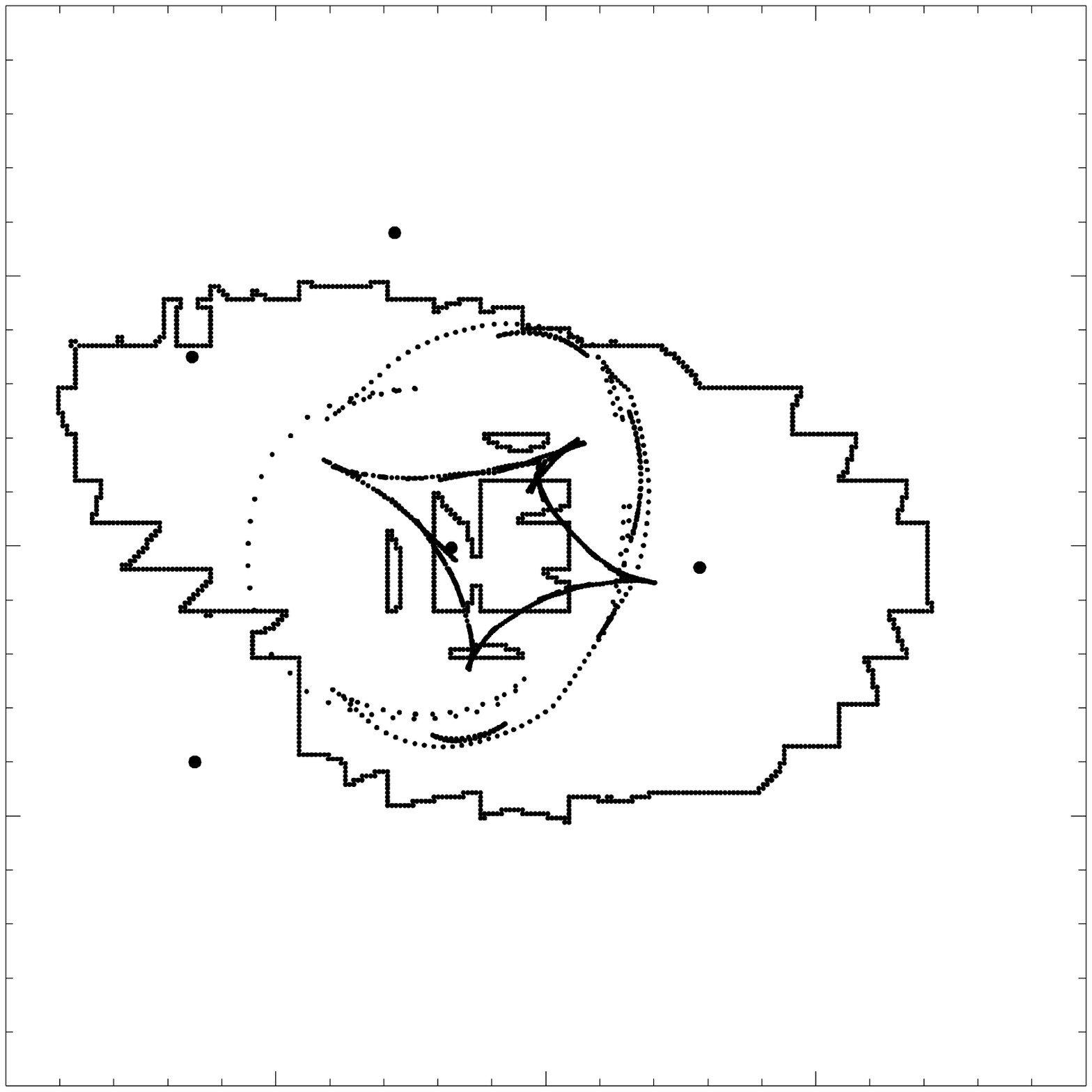}{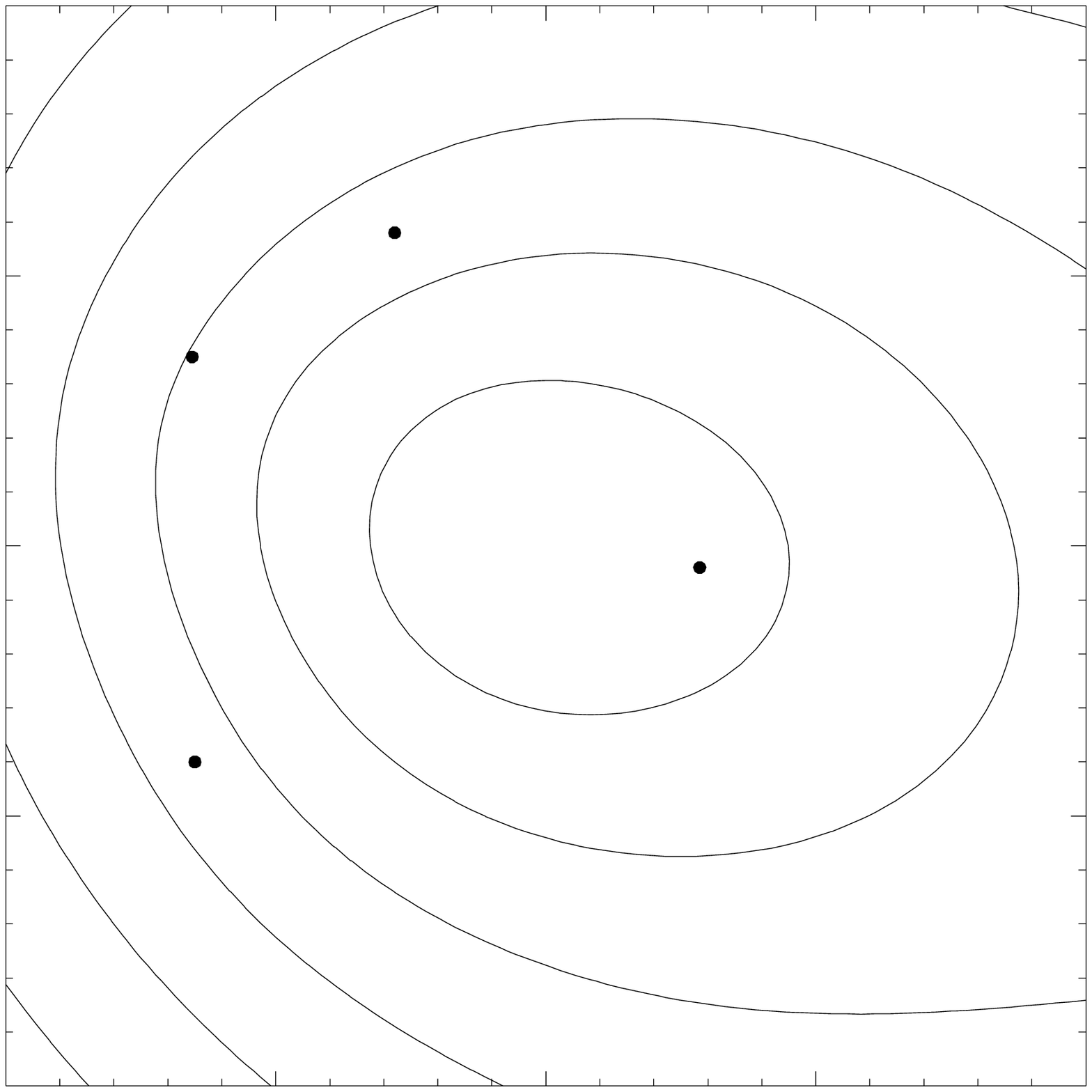}{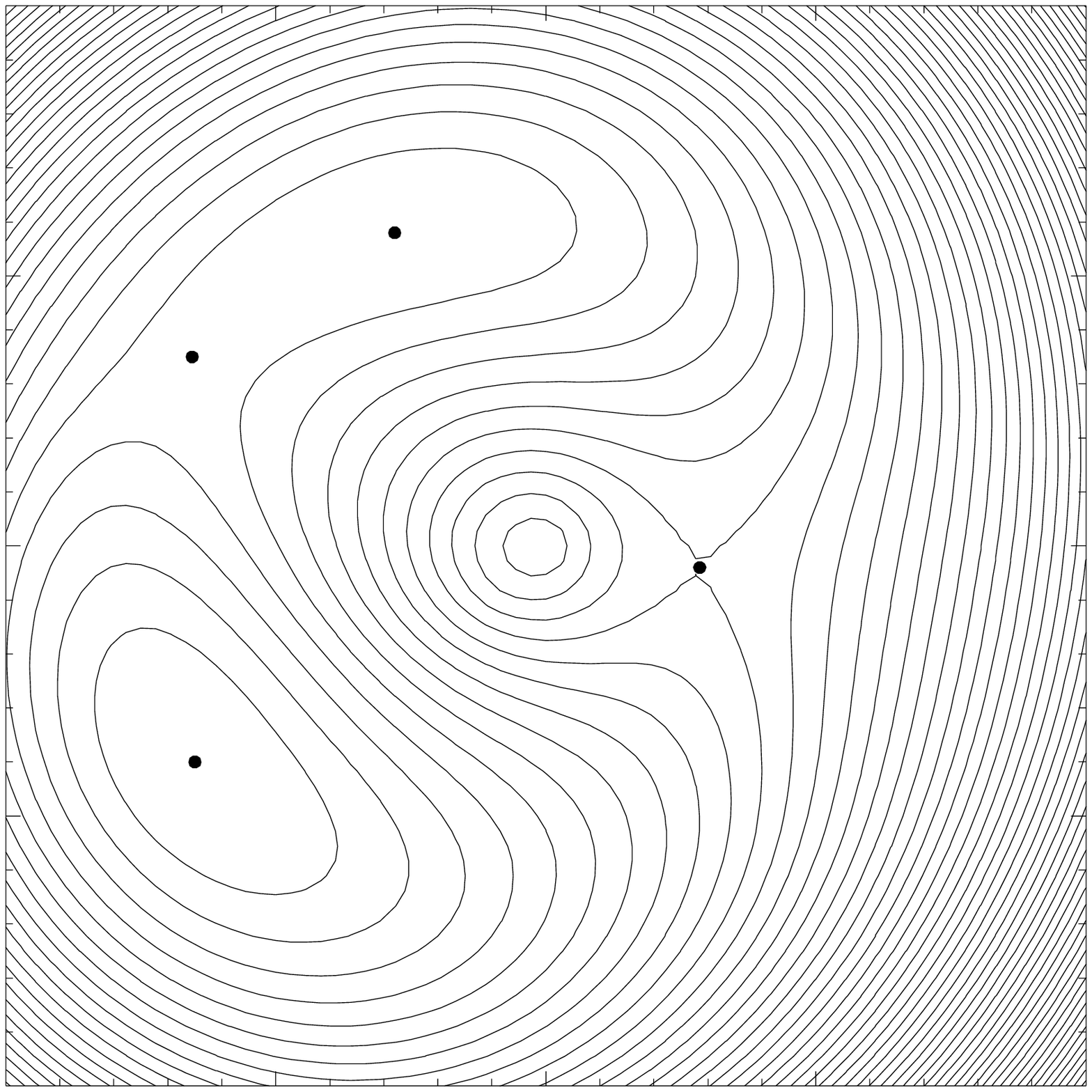}
\caption
{Same as Figure~\ref{four_1115}, but for B1608+656.
\label{four_1608}}
\end{figure}

\begin{figure}
\plottwo{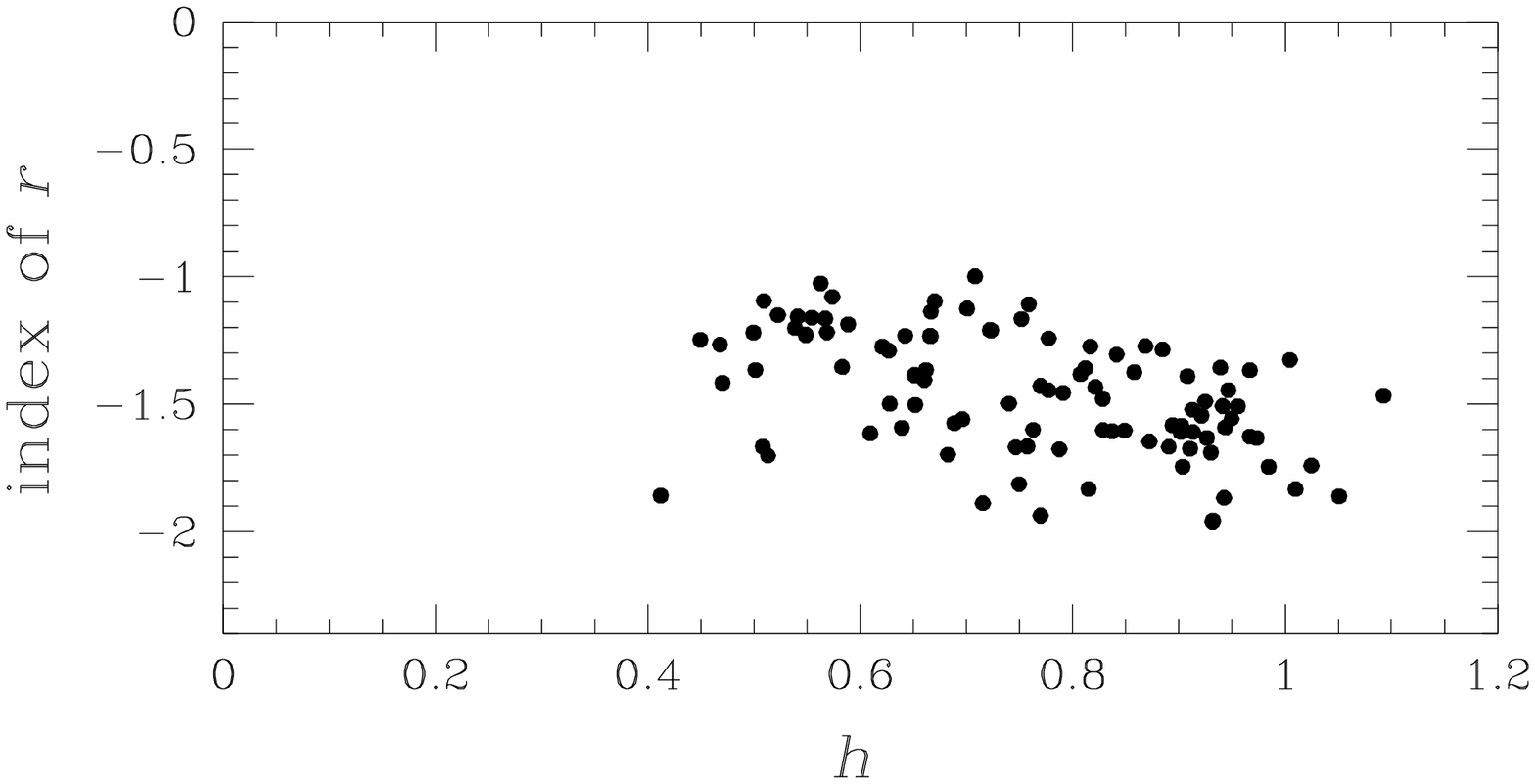}{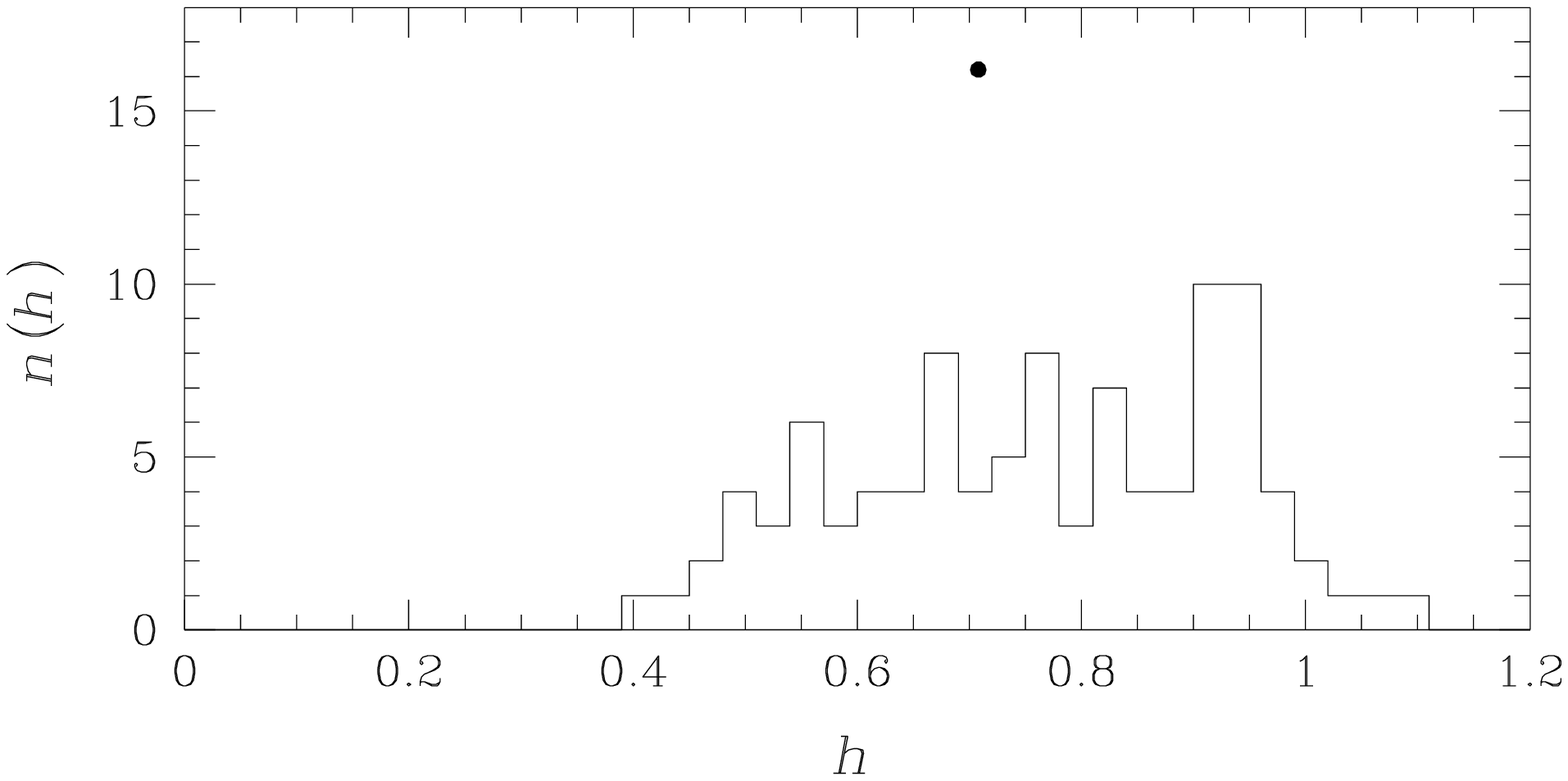}
\caption
{B1608+656: (a) Slope of the projected density profile, ${\rm ind}(r)$,
plotted against the derived $h$ value; (b) Probability distribution of derived 
$h$ values. Solid dot marks the location of the `most isothermal' galaxy.
\label{two_1608}}
\end{figure}

The light distribution of the lensing system is rather messy, possibly
representing a merging/interacting galaxy pair, therefore inversion
symmetry was not used in the following reconstructions. 
Figures~\ref{four_1608} and \ref{two_1608} are the same as 
Figures~\ref{four_1115} and \ref{two_1115}, but for B1608.
The range 50 to 100$\rm\,km\,sec^{-1}\,Mpc^{-1}$ in Figure~\ref{two_1608}(b)
encompasses about 90$\%$ of the reconstructions.

\subsection{Combined p(h) plot}
\begin{figure}
\plotone{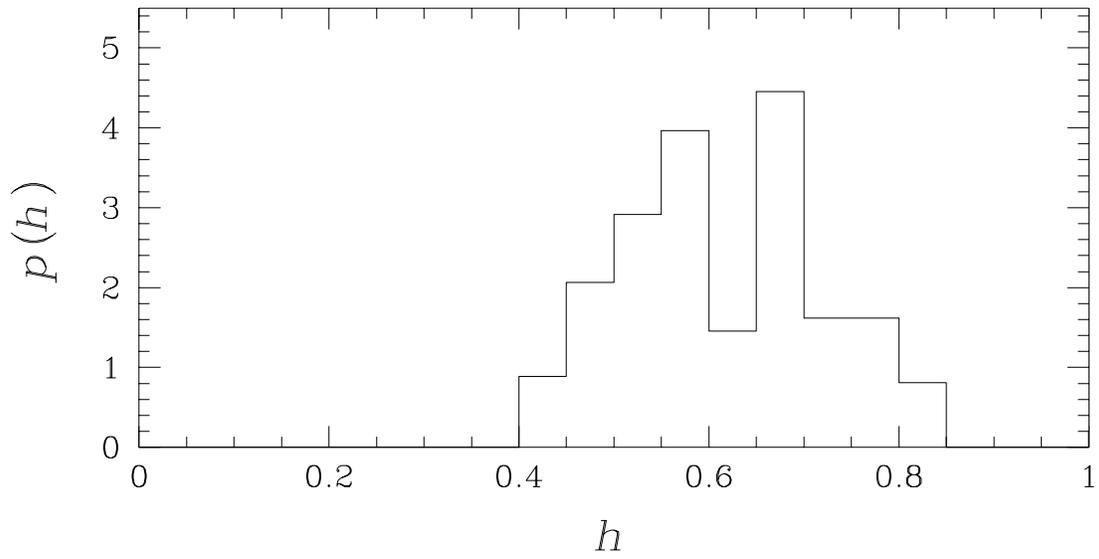}
\caption
{Combined $H_0$ probability distribution derived from the 
non-parametric reconstruction of galaxy lenses in PG1115 and B1608.
$90\%$ of all points lie within the range 43--79$\rm\,km\,sec^{-1}\,Mpc^{-1}$; 
the median of the distribution is 61$\rm\,km\,sec^{-1}\,Mpc^{-1}$.
\label{phreal_fig}}
\end{figure}

Just like in the case of the blind test we now multiply probability
distributions from PG1115 and B1608 to get the combined distribution,
Figure~\ref{phreal_fig}. $90\%$ of all points lie within the range
43--79$\rm\,km\,sec^{-1}\,Mpc^{-1}$, while the median of the
distribution is 61$\rm\,km\,sec^{-1}\,Mpc^{-1}$. Note that the
errorbars obtained using our method are substantially larger than what
is usually quoted in other studies.  We ascribe this increase to the
more systematic sampling of the whole image-defined model space
unrestricted by the confines of parametric models.

\section{Discussion and Conclusions}\label{summary}

Multiply-imaged QSO systems provide us with an elegant way of
measuring $H_0$, and a lot of observational and modeling effort has
been invested in this enterprise.  As the quality of the observational
data improves, most of the uncertainty in $H_0$ is contributed by the
mass distribution in the lens. How to treat this problem is a matter
of some debate. Should one use a single, physically motivated mass
model, or should one approach the problem with no preconceptions about
the galaxy form?

In general, ad hoc restrictions on the
allowed mass models translate into a too optimistic and probably
biased estimated distribution of $H_0$'s. To avoid this trap one has
to allow as much freedom for the lens models as possible.  On the
other hand, to yield a useful estimate of $H_0$ one has to restrict
the amount of freedom allowed for the models using physically
motivated criteria. Ideally one wants to balance these two opposing
tendencies and impose just the correct quantity and quality of model
constraints. Based on our experience from the present work we conclude
that parametric, or any other approach that severely restricts the freedom
of the galaxy-lens, has over-constrained their models and thus
ended up with unrealistically small errorbars, and biased $H_0$'s. As a
result different models of the same systems can yield discrepant results.
For example, Romanowski \& Kochanek (1998) use dynamical methods to model 
the galaxy in Q0957, and further constrain the galaxy to be similar
to nearby ellipticals; they quote $61^{+13}_{-15}$ at $2\sigma$ level. 
Bernstein \& Fischer (1999) analyzed the same system but used a {\it range}
of astrophysically reasonable parametric models. Their estimate,
$77^{+29}_{-24}$, also at $2\sigma$ level, does not agree with that of
Romanowski \& Kochanek. Our approach is different in that it does not
presuppose a galaxy shape, but instead allows us to impose as many or as 
few constraints as is deemed appropriate.

The most unrestricted models would be constrained solely by what we call the 
primary constraints, i.e., image observables. By definition these would yield 
unbiased estimates of $H_0$ {\it based on lensing data alone}. We chose to 
go somewhat beyond this and apply what we call secondary constraints, which 
describe realistic galaxies in most general terms. The derived $H_0$ 
distributions are narrower; the price we pay is a small amount of bias. 
It can be argued that we are still too generous with our mass models, i.e.
other galaxy characteristics can be safely assumed, and hence
tighter constraints can be applied to the models without sacrificing the 
unbiased nature of results. This avenue can be taken in the future work if
additional constraints become available. 

A potential source of additional modeling constraints are optical
rings, lensed images of the QSO galaxy host, which are seen in some cases,
for example, in PG1115 and B0218. The orientations and elongations
of individual images of QSO host galaxy can be used as linear inequality 
constraints to narrow down the range of possible galaxy mass distributions.

If two or more sources with known redshifts were lensed by the same 
foreground galaxy, these could be used to break the mass sheet degeneracy
and thus further constrain the galaxy. However in practice cases
of two sources at different redshifts lensed by the same galaxy are expected
to be very rare because of the small galaxy cross-sections.

Probably the most promising potential constraint is based on the relation 
between the slope of the projected density profile around the images and the 
derived $H_0$. If the slope can be estimated by means other than lensing, or 
at least a limit placed on its value as we did in Section~\ref{PG1115} using
the observed slope of the light distribution, then
$H_0$ can be constrained much better compared to what is currently possible.

With the two systems used in the present work, PG1115+080 and B1608+656, 
and implementing primary constraints of image properties and secondary 
constraints describing a few general properties of lensing galaxies, we
conclude that $H_0$ is between 43 and 79$\rm\,km\,sec^{-1}\,Mpc^{-1}$ at $90\%$ 
confidence level, with the best estimate being 61$\rm\,km\,sec^{-1}\,Mpc^{-1}$.

\end{document}